\documentclass[reprint,twocolumn,superscriptaddress,letter,floatfix]{revtex4-2}
\usepackage[english]{babel}
\usepackage[utf8]{inputenc}
\usepackage[T1]{fontenc}
\usepackage{amsmath,amssymb,braket,slashed,mathrsfs}
\usepackage{pifont}
\usepackage{hyperref}
\usepackage{color,hyperref}
\usepackage{physics}
\usepackage{cleveref}
\usepackage{amsmath}
\usepackage{graphicx}
\usepackage{booktabs}
\usepackage{multirow}
\usepackage{subfigure}
\usepackage{float}
\usepackage{epstopdf}

\newcommand{\beq}{\begin{eqnarray}}
\newcommand{\eeq}{\end{eqnarray}}
\newcommand{\be}{\begin{equation}}
\newcommand{\ee}{\end{equation}}
\newcommand{\ba}{\begin{eqnarray}}
\newcommand{\ea}{\end{eqnarray}}

\newcommand{\bit}{\begin{itemize}}
\newcommand{\eit}{\end{itemize}}

\newcommand{\innovation}{Collaborative Innovation Center of Quantum Matter, Beijing 100871, China}
\newcommand{\chep}{Center for High Energy Physics, Peking University, Beijing 100871, China}
\newcommand{\pkuphy}{School of Physics, Peking University, Beijing 100871,
China}

\begin{document}

\title{Lattice calculation of $\chi_{c0} \rightarrow 2\gamma$ decay width}

\author{Zuoheng Zou}\affiliation{\pkuphy}
\author{Yu Meng}\affiliation{\pkuphy}\affiliation{\chep}
\author{Chuan Liu}\affiliation{\pkuphy}\affiliation{\chep}\affiliation{\innovation}

\date{\today}

\begin{abstract}
We perform a lattice QCD calculation of the $\chi_{c0} \rightarrow 2\gamma$ decay width using a model-independent method which does not require a momentum extrapolation of the corresponding off-shell form factors. The simulation is performed on ensembles of $N_f=2$ twisted mass lattice QCD gauge configurations with three different lattice spacings. After a continuum extrapolation, the decay width is obtained to be $\Gamma_{\gamma\gamma}(\chi_{c0})=3.65(83)_{\mathrm{stat}}(21)_{\mathrm{lat.syst}}(66)_{\mathrm{syst}}\, \textrm{keV}$.
 Albeit this large statistical error, our result is compatible with the experimental
 results within 1.3$\sigma$.
 Potential improvements of the lattice calculation in the future are also discussed.


\end{abstract}
\maketitle

\section{Introduction}
\label{sec:intro}

Charmonium physics lives in an energy regime where both perturbative and nonperturbative features of quantum chromodynamics (QCD) intertwine. Notably, Charmonium decay has played an important role in establishing the asymptotic freedom of QCD and served as a clean platform to probe the interplay between perturbative and nonperturbative dynamics. In particular, the two photon annihilation rates of charmonium are extremely helpful for the understanding of quark-antiquark interaction and the decay mechanisms~\cite{huang96,hwang10}.

 In this paper, we study the two-photon decay width of $\chi_{c0}$,
 which has been extensively studied from both experimental and theoretical sides. On the experimental side, using the decay of $\psi(3686)\rightarrow\gamma\chi_{c0},\chi_{c0}\rightarrow\gamma\gamma$, both CLEO-c and BESIII collaborations reported results of the two-photon decay width $\Gamma_{\gamma\gamma}(\chi_{c0})$~\cite{cleo, bes}:\\

 \be
 \begin{aligned}
 \Gamma^{CLEO-c}_{\gamma\gamma}(\chi_{c0}) &=2.36(35)_{\mathrm{stat}}(22)_{\mathrm{syst}}\, \textrm{keV}
 \\
 \Gamma^{BESIII}_{\gamma\gamma}(\chi_{c0}) &=2.03(8)_{\mathrm{stat}}(14)_{\mathrm{syst}}\, \textrm{keV}
 \\
 \Gamma^{PDG}_{\gamma\gamma}(\chi_{c0}) &=2.20(22) \textrm{keV}
 \end{aligned}
 \ee
 where the first line from CLEO-c, the second from BESIII
 and the last line being the PDG quoted value with combined errors.
 It is expected that more accurate results for these decay width
 will become available in the near future.

 On the theoretical side, it is fair to say that the situation is far from satisfactory.
 Theoretical results for the decay rate have been obtained using a non-relativistic approximation~\cite{appel,Barnes},
 potential model~\cite{Gupta}, relativistic quark model~\cite{ebert,godfrey,Bodwin,Munz}, non-relativistic QCD (NRQCD) factorization~\cite{BARBIERI1976183, BARBIERI80,BARBIERI81, MA2002233,Brambilla,SCHULER,sang16}, effective Lagrangian~\cite{lansberg09},
 Dyson-Schwinger equations (DSEs)~\cite{chen17}, as well as quenched~\cite{dudek06}
 and unquenched lattice calculations~\cite{Chen_2020}.
 These results are listed in Table~\ref{table:theory},
 which scatter quite a lot although all fall in the right ballpark.
 Note that within the framework of NRQCD, the leading-order (LO) prediction is close to the experimental measurements, but this process is extremely sensitive to high-order QCD radiative corrections and relativistic corrections.
 %in Ref.~\cite{SCHULER}, the authors quoted a value of $\Gamma_{\gamma\gamma}(\chi_{c0})=2.5$keV
 %in the leading order of NRQCD. However, in a more recent higher order computation presented in Ref.~\cite{sang16},
 %the authors found large corrections from higher orders and did not quote a direct result for $\Gamma_{\gamma\gamma}(\chi_{c0})$.
 Therefore, only the LO predictions are listed in Table~\ref{table:theory} .
 %\textcolor{blue}{Please check if this statement is true or not?}

\begin{table}[!htb]
\begin{ruledtabular}
\begin{tabular}{cccc}
\multicolumn{4}{c}{Theoretical computations for $\Gamma_{\gamma\gamma}(\chi_{c0})$(keV)}\\
\hline
Huang~\cite{huang96}     &    3.72~$\pm$~1.10      & Barbieri~\cite{BARBIERI1976183} &   3.5   \\
Barnes~\cite{Barnes}     &   1.56                  & Schuler~\cite{SCHULER}             &   2.50 \\
Gupta~\cite{Gupta}       &   6.38                  & Lanseberg~\cite{lansberg09}      &  5.00   \\
Ebert~\cite{ebert}       &  2.90                   & Chen~\cite{chen17}               &  2.06-2.39   \\
Godfrey~\cite{godfrey}   &  1.29                   &  Crater~\cite{Crater}            &   3.34-3.96  \\
Bodwin~\cite{Bodwin}     &  6.70~$\pm~2.80$        & Wang~\cite{WANG}                 &   3.78    \\
M\"{u}nz~\cite{Munz}     &  1.39~$\pm$~0.16        &  Laverty~\cite{laverty2011gaga}  &   1.99-2.10    \\
\hline
Dudek~\cite{dudek06}     & $2.41(58)_{\mathrm{stat}}(86)_{\mathrm{syst}}$
&  CLQCD~\cite{Chen_2020}  &   $0.93(19)_{\mathrm{stat}}$
\end{tabular}
\end{ruledtabular}
\label{table:theory}
\caption{Some theoretical predictions for  $\Gamma_{\gamma\gamma}(\chi_{c0})$.}
\end{table}

 In the last line of Table~\ref{table:theory}, we list two existing lattice QCD results so far.
 The first one from Dudek {\em et al} is a quenched lattice computation on a single lattice spacing~\cite{dudek06}.
 The systematic error they quote mainly come from quenching. The second one from CLQCD is an unquenched
 study using $N_f=2$ twisted mass fermions at two distinct lattice spacings. The authors found
 that the lattice artifacts are substantial and only quoted results from a finite lattice
 spacing, without an error estimate of the finite lattice spacing errors.
 The number quoted in Table~~\ref{table:theory} is the result from the finer lattice  spacing~\cite{Chen_2020}.
 Therefore, in both lattice studies, systematic effects such as finite lattice errors
 are not fully investigated which was found to be large in the second study~\cite{Chen_2020}.
 Obviously, in order to fully compare with the upcoming experiments,
 one needs to work in a theoretical framework that allows an improvable error control
 and in this respect, lattice computation obviously has an advantage over other phenomenological methods
 listed in Table~\ref{table:theory}.

 In this paper, we try to improve on the existing lattice computation of $\Gamma_{\gamma\gamma}(\chi_{c0})$
 in two major aspects: First, in previous lattice studies,
 many systematic effects are not yet fully taken into account,
 the most important of which being the finite lattice spacing effect, which has been observed in Ref.~\cite{Chen_2020}.
 Second, one normally computed the off-shell form factors at various discrete photon
 virtualities. In order to obtain the physical decay width, an extrapolation of these
 results are required, introducing a model-dependent systematic error.

 In this work, we have made the following improvements:
 First, to attack the lattice artifacts, we perform our calculation on ensembles with three different lattice spacings,
 allowing us to perform a reliable continuum extrapolation.
 Second, we adopt a novel method to extract the on-shell form factor directly,
 by-passing the conventional momentum extrapolation and therefore
 avoids the corresponding model-dependent extrapolation errors.
 We have also taken the excited-state contamination into consideration,
 further improving our results on the physical form factor.
 Similar procedures has been successfully utilized to two-photon decay of $\eta_c$~\cite{meng2021firstprinciple}.
 We hope that these improvements could also shed some light on the two-photon decay of $\chi_{c0}$.

 This paper is organized as follows. In Sect.~\ref{sec:method}, the methodology for extracting on-shell form factor is introduced. The Sect.~\ref{sec:simulation results} is divided into several parts: In Sect.~\ref{sec:lattice setup},
 the information of the configurations and operators used in this work are introduced.
 In Sect.~\ref{sec:mass}, the mass spectrum of $\chi_{c0}$ is presented.
 In Sect.~\ref{sec:Z}, we give the renormalization factor and the spectrum weight factor.
 In Sect.~\ref{sec:form factor}, numerical results of the form factor in three different lattice spacings are presented.
 Then in Sect.~\ref{sec:final result}, extrapolation of the results to continuum is performed, yielding
 our final result for the decay rate. We also compare our result with both experimental and theoretical results.
 The main sources of error in our work are discussed and possible solutions in the future are proposed.
%write sect first
%
\section{Methodology}
\label{sec:method}
 In this section, we outline the methodology for the calculation of two-photon decay width of $\chi_{c0}$.
 In the traditional approach~\cite{Ji}, using the Lehmann-Symanzik-Zimmermann (LSZ) reduction formula and
 integrating out the QED part to $\mathcal{O}(\alpha_{em})$,
 the amplitude for two-photon decay of charmonium
 can be obtained as follows~\cite{dudek06},
\be
\begin{aligned}
\langle\gamma&\gamma|M(p_f)\rangle\sim e^2\epsilon^*_\mu\epsilon^*_\nu\int dt_i e^{- \omega_1 (t_i -t)}\int d^3 \vec{x}\, e^{-i\vec{p_f}.\vec{x}}\\
&\int d^3 \vec{y}\, e^{i\vec{q_2}.\vec{y}}\langle 0 | T\Big\{
\varphi_{M}(\vec{x}, t_f)
J_\nu(\vec{y}, t) J_\mu(\vec{0}, t_i)\Big\}|0\rangle
\label{eq:LSZ}
\end{aligned}
\ee
 where $\varphi_{M}(\vec{x},t_f)$ is an appropriate composite operator
 which creates a desired meson $M$ (in our case, the $\chi_{c0}$ meson) from the QCD vacuum;
 $\epsilon_{\mu},\epsilon_{\nu}$ are the polarization four-vectors for the two final photons;
 $J_\mu=\sum_qe_q\,\bar{q}\gamma_\mu q$ ($e_q=2/3,-1/3,-1/3,2/3$ for $q=u,d,s,c$)
 is the electromagnetic current operator due to the quarks, with $e$ being the elementary charge unit.
 In this work, we only consider the connected contributions arising from the charm quark current.
 Disconnected contributions are neglected. These contributions are extremely costly for
 lattice computations and are assumed to be small in charmonium physics~\cite{dudek06,ukqcd, taro}.
 Then, the matrix element in Eq.~(\ref{eq:LSZ}) relevant for $\chi_{c0}$ decay
 can be parameterized in terms of the form factor $G(Q_1^2,Q_2^2)$ as,
\beq
\label{eq:form}
&&\langle\gamma(q_1) \gamma(q_2) |M(p_f)\rangle \nonumber\\
\!\!\!\!\!\!\!\!\!\!\!\!\!\!\!\!&=&\tfrac{2}{m_{\chi}}(\tfrac{2}{3} e)^2 G(Q_1^2,Q_2^2)\left[\epsilon_1\cdot\epsilon_2q_1\cdot{q_2}-\epsilon_2\cdot{q_1}\epsilon_1\cdot{q_2}\right]
\eeq
 where $q_1,q_2$ are the two four-momenta of the final photons while $Q^2_1=-q^2_1,Q^2_2=-q^2_2$
 are the virtualities of the two photons. The mass of $\chi_{c0}$ is denoted as $m_\chi$ and
 the polarization vectors of the two photons are given by $\epsilon_1$ and $\epsilon_2$.
 The physical decay width is related to the on-shell form factor
 which is obtained by a momentum extrapolation towards the physical point: $Q^2_1=Q^2_2=0$.
 Thus, in this conventional approach, in order to have a better control on the extrapolation,
 one needs to compute the matrix element at various different non-physical virtuality combinations,
 thereby also introducing extra computational costs.
 The extrapolation itself also brings about model-dependent systematic errors.
 In the new approach introduced in this work,
 we adopt a method that requires no off-shell form factor calculations at all
 and therefore by-passing the model-dependent extrapolation in photon virtualities.
 The method has been successfully utilized in two-photon decays of $\eta_c$~\cite{meng2021firstprinciple}.
 We now briefly outline the major steps for the case of $\chi_{c0}$ below.

% \textcolor{blue}{This should be an $e_c$, not $q_c$, according to the expression you gave above!}\\
% \textcolor{blue}{I re-formulate the following paragraph. The original one was not clear enough. For example,
% the four-vectors are messed up...}\\
 One first relates the on-shell decay amplitude of $\chi_c\rightarrow 2\gamma$ to
 an infinite-volume hadronic tensor $\mathcal{F}_{\mu\nu}(p)$ which is the
 Fourier transform of the real-space tensor $\mathcal{H}_{\mu\nu}(t,\vec{x})$ in
 continuum Euclidean space,
 \begin{equation}
 \begin{aligned}
  &\mathcal{F}_{\mu\nu}(p) =\int dt e^{m_\chi t/2}\int d^3 \vec{x} e^{-i\vec{p}\cdot\vec{x}}\mathcal{H}_{\mu\nu}(t,\vec{x})\;,\\
&\mathcal{H}_{\mu\nu}(t,\vec{x}) =\bra{0}T{J_\mu(x)J_\nu(0)}\ket{\chi_{c0}(k)}\;,
\end{aligned}
\end{equation}
 where  we have chosen the rest-frame of the $\chi_{c0}$ meson so that $k=(im_\chi, \vec{0})$.
 Note that we have fixed the four-momentum for one of the final photons to be $p=(im_\chi/2,\vec{p})$
 with $|\vec{p}|=m_\chi/2$, making it on-shell explicitly and energy-momentum conservation then
 guarantees the other photon with four-momentum $p^\prime$ is also on-shell.
 With this choice, the on-shell decay amplitude may be written as,
 \begin{equation}
  M=e^2\epsilon^*_\mu(p,\lambda)\epsilon^*_\nu(p^\prime,\lambda^\prime)\mathcal{F}_{\mu\nu}(p)
 \end{equation}
 According to the quantum number of $\chi_{c0}$,
 the hadronic tensor can be parameterized as (repeated indices are summed),
 \begin{equation}
 \mathcal{F}_{\mu\nu}(p)=\epsilon_{ij\mu\alpha}\epsilon_{ij\nu\beta}p_\alpha k_\beta
 F_{\chi_{c0}\gamma\gamma}.
 \end{equation}
 The approach to extract the on-shell form factor $F_{\chi_{c0}\gamma\gamma}$ here
 is also slightly different from the conventional one.
 By further multiplying the Lorentz structure factor in the above equation, the hadronic tensor can be
 contracted to a scalar including only the form factor $F_{\chi_{c0}\gamma\gamma}$ with a constant factor.
 Then the form factor can be derived by dividing the coefficient as follows,
\begin{equation}
\begin{aligned}
F_{\chi_{c0}\gamma\gamma}&=\frac{\epsilon_{ij\mu\alpha}\epsilon_{ij\nu\beta}p_\alpha k_\beta\mathcal{F}_{\mu\nu}(p)}{\epsilon_{ij\mu\alpha}\epsilon_{ij\nu\beta}p_\alpha k_\beta\epsilon_{i'j'\mu\alpha'}\epsilon_{i'j'\nu\beta'}p_{\alpha'} k_{\beta'}}\\
&=-\frac{1}{8m_{\chi}|\vec{p}|^2}\int dte^{m_\chi t/2}\\
&\times\int d^3\vec{x}e^{-i\vec{p}\cdot\vec{x}}\epsilon_{ij\mu\alpha}\epsilon_{ij\nu0}\frac{\partial\mathcal{H}_{\mu\nu}(x)}{\partial x_{\alpha}}\\
\end{aligned}
\end{equation}
 Until now, all derivations are in the continuum Euclidean space.
 We now utilize the spatial isotropy symmetry to average over the spatial direction of $\vec{p}$,
\begin{equation}
\begin{aligned}
e^{-i\vec{p}\cdot\vec{x}}&\rightarrow \frac{1}{4\pi}\int d\Omega_{\vec{p}} e^{-i\vec{p}\cdot\vec{x}}=\frac{\sin(|\vec{p}||\vec{x}|)}{|\vec{p}||\vec{x}|}\equiv j_0(|\vec{p}||\vec{x}|)\\
&\frac{d}{dz}(j_0(z)) =-\left(\frac{\sin z}{z^2}-\frac{\cos z}{z}\right) \equiv -j_1(z),
\end{aligned}
\end{equation}
 where $j_n(x)$ are the spherical Bessel functions. Finally the scalar from factor is expressed as
\begin{equation}
\begin{aligned}
\label{eq:Fchi}
F_{\chi_{c0}\gamma\gamma}&=\frac{1}{8m_{\chi}}\int dte^{m_\chi t/2}\int d^3\vec{x}\\
&\!\!\!\!\!\!\!\!\!\times\left[\frac{j_1(|\vec{p}||\vec{x}|)}{|\vec{p}||\vec{x}|}(x_i\mathcal{H}_{0i}+x_i\mathcal{H}_{i0})
+\frac{j_0(|\vec{p}||\vec{x}|)}{|\vec{p}|}2\mathcal{H}_{ii}\right]\\
\end{aligned}
\end{equation}
 where $i=1,2,3$  take spatial indices and are assumed to be summed over.

To obtain the hadronic tensor $\mathcal{H}_{\mu\nu}(t,\vec{x})$ in Eq.~(\ref{eq:Fchi}),
we utilize the variational method to find the optimal interpolation operators to create the $\chi_{c0}$ meson state~\cite{shultz15}. The physical decay width of $\chi_{c0}$ is given by
\begin{equation}
\Gamma_{\gamma\gamma}(\chi_{c0}) = \alpha^2\pi m_{\chi_{c0}}^3F_{\chi_{c0}\gamma\gamma}^2\;.
\label{eq:gamma}
\end{equation}
  Therefore, one only needs to compute the Euclidean correlation functions $\mathcal{H}_{0i}$
  and $\mathcal{H}_{ii}$  that are directly relevant for the on-shell amplitude
  and substitute the results into Eq.~(\ref{eq:Fchi}) to arrive
  at the physical decay width $\Gamma_{\gamma\gamma}(\chi_{c0})$ in Eq.~(\ref{eq:gamma}).
  This completely avoids the on-shell extrapolation process in the conventional lattice approach.

\section{Simulation results}
\label{sec:simulation results}

\subsection{Lattice setup}
\label{sec:lattice setup}
 We utilize three $N_f=2$-flavor twisted mass gauge field ensembles generated by the Extended Twisted Mass Collaboration(ETMC) with lattice spacing $a \simeq 0.0667,0.085,0.098$ fm, respectively.
 The  parameters of these ensembles are presented in Table.~\ref{table:cfgs}.
 The valence charm quark mass parameter $\mu_c$ is tuned so that
the mass of the $\eta_c$ meson for each ensemble reproduces its
correct physical value.
 For more details, we refer the reader to Ref.~\cite{BOUCAUD2008695,tw2010},
\begin{table}[!htb]
\begin{ruledtabular}
\begin{tabular}{ccccccc}
\textrm{Ensemble} & $a$ (fm) & $L^3\times T$  & $N_{\textrm{conf}}$
& $a\mu_l$ & $m_\pi$ (MeV) & $t_h$  \\
%\multicolumn{1}{c}{\textrm{Three}}& \\
\hline
Ens.I            &  0.067(2)   & $32^3\times 64$ & $179$ & 0.003 & 300  &10-20 \\
Ens.II           &  0.085(3)   & $24^3\times 48$ & $200$ & 0.004 & 315  & 10-15     \\
Ens.III          &  0.098(3)   & $24^3\times 48$ & $216$ & 0.006 & 365   & 10-15       \\
\end{tabular}
\end{ruledtabular}
\label{table:cfgs}
\caption{\label{table:cfgs}%
 From left to right, we list the ensemble name, the lattice spacing $a$,
 the spatial and temporal lattice size $L$ and $T$, the number of the measurements
of the correlation function for each ensemble $N_{\textrm{conf}}\times T$, the light quark mass $a\mu_l$, the pion mass $m_\pi$ and the range of the time separation $t_h$ between $\chi_{c0}$ and photon.}
\end{table}

 Before getting into the simulation details,
 there remains one subtlety to clarify that is related to the twisted mass fermion.
 Since the twisted mass action breaks parity $\mathcal{P}$ by $\mathcal{O}(a^2)$ effects, the basis operator $\mathcal{O}_1=\bar{c}c$ for $\chi_{c0}$ would unfortunately mix with $\mathcal{O}_2=\bar{c}\gamma^5c$
 which has the opposite parity. This mixing implies that a specific combination of these operators will be relevant to create a physcial scalar charmonium in the twisted mass action~\cite{shultz15}
\begin{equation}
\mathcal{O}^\dagger_{\chi_{c0}} = v^{\chi}_{1}\mathcal{O}^\dagger_1 + v^{\chi}_{2}\mathcal{O}^\dagger_{2}
\end{equation}
 The two-point correlation function $C_{\chi_{c0}}(t)=\bra{0}\mathcal{O}_{\chi_{c0}}(t)\mathcal{O}^\dagger_{\chi_{c0}}(0)\ket{0}$
 can be derived by multiplying the corresponding coefficients with the basis correlation functions $C'_{ij}=\bra{0}\mathcal{O}_i(t)\mathcal{O}^\dagger_{j}(0)\ket{0}(i,j=1,2)$.
 Therefore, after choosing a time slice $t_0$,
 one could disentangle the mixing of the two operators by solving
 a generalized eigenvalue problem (so-called GEVP procedure):
\begin{equation}
\label{eq:gevp}
\begin{aligned}
&\left (
\begin{matrix}
C'_{11}(t) & C'_{12}(t) \\\
C'_{21}(t) & C'_{22}(t)
\end{matrix}
\right )
\left (
\begin{matrix}
v^{\chi}_{1} & v^{\eta}_{1} \\\
v^{\chi}_{2} & v^{\eta}_{2}
\end{matrix}
\right )\\
=&\left (
\begin{matrix}
\lambda_1 & 0 \\\
0 & \lambda_2
\end{matrix}
\right )
\left (
\begin{matrix}
C'_{11}(t_0) & C'_{12}(t_0) \\\
C'_{21}(t_0) & C'_{22}(t_0)
\end{matrix}
\right )
\left (
\begin{matrix}
v^{\chi}_{1} & v^{\eta}_{1} \\\
v^{\chi}_{2} & v^{\eta}_{2}
\end{matrix}
\right )
\end{aligned}
\end{equation}
where the generalized eigenvalues $\lambda_i$ behave like $e^{-E_i(t-t_0)}$ at large time separation.
 In practice, we fix $t_0=1$ and solve Eq.~(\ref{eq:gevp}) on each time-slice independently
 and use them to reconstruct the three-point correlation functions.

\subsection{Mass spectrum for $\chi_{c0}$}
\label{sec:mass}
\begin{figure}[!htb]
\begin{minipage}{0.95\linewidth}
  %\centerline
  {\includegraphics[width=0.95\linewidth]{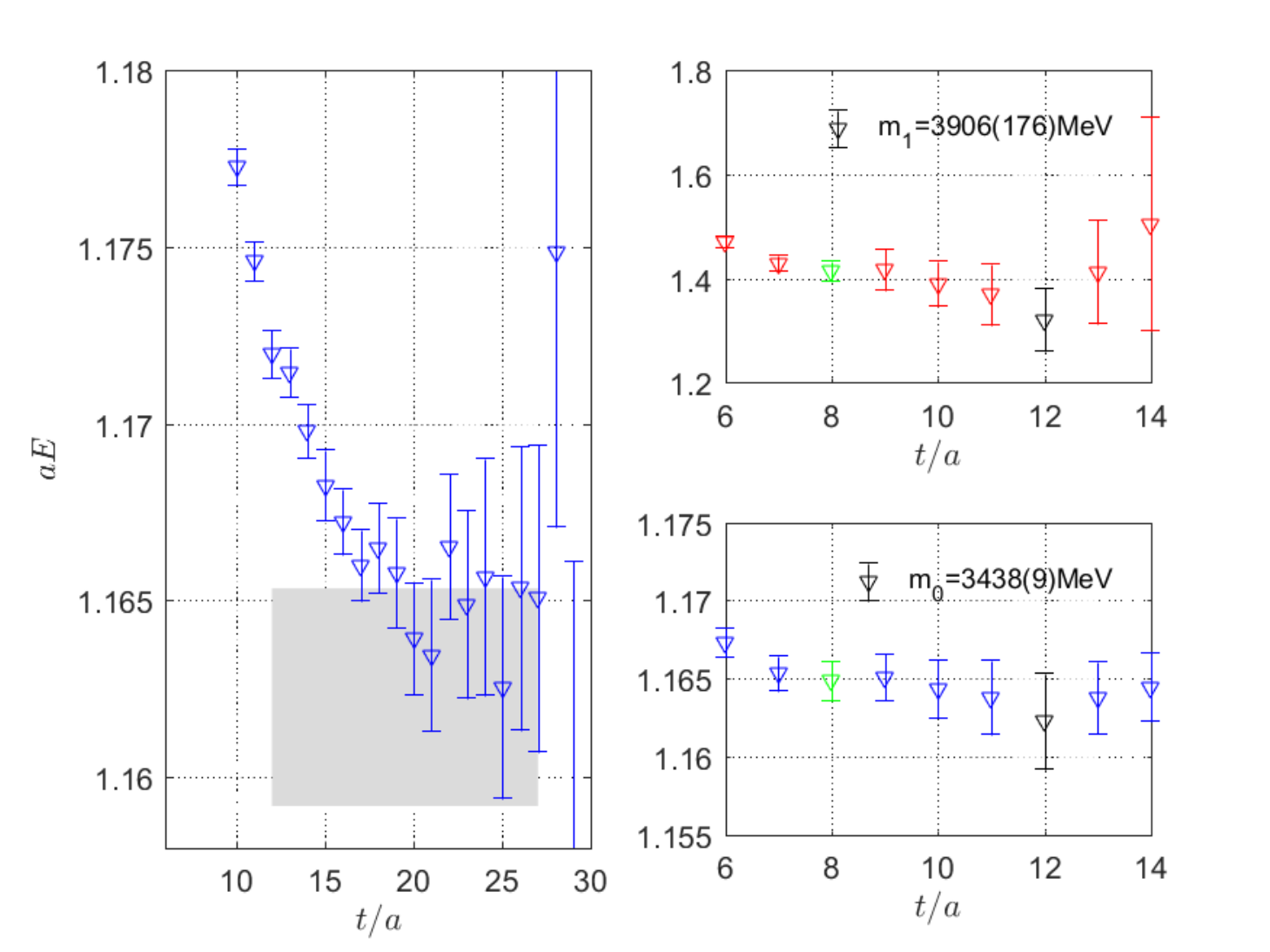}}
  \center (a) Ens.I
\end{minipage}\\
\begin{minipage}{0.95\linewidth}
  %\centerline
  {\includegraphics[width=0.95\linewidth]{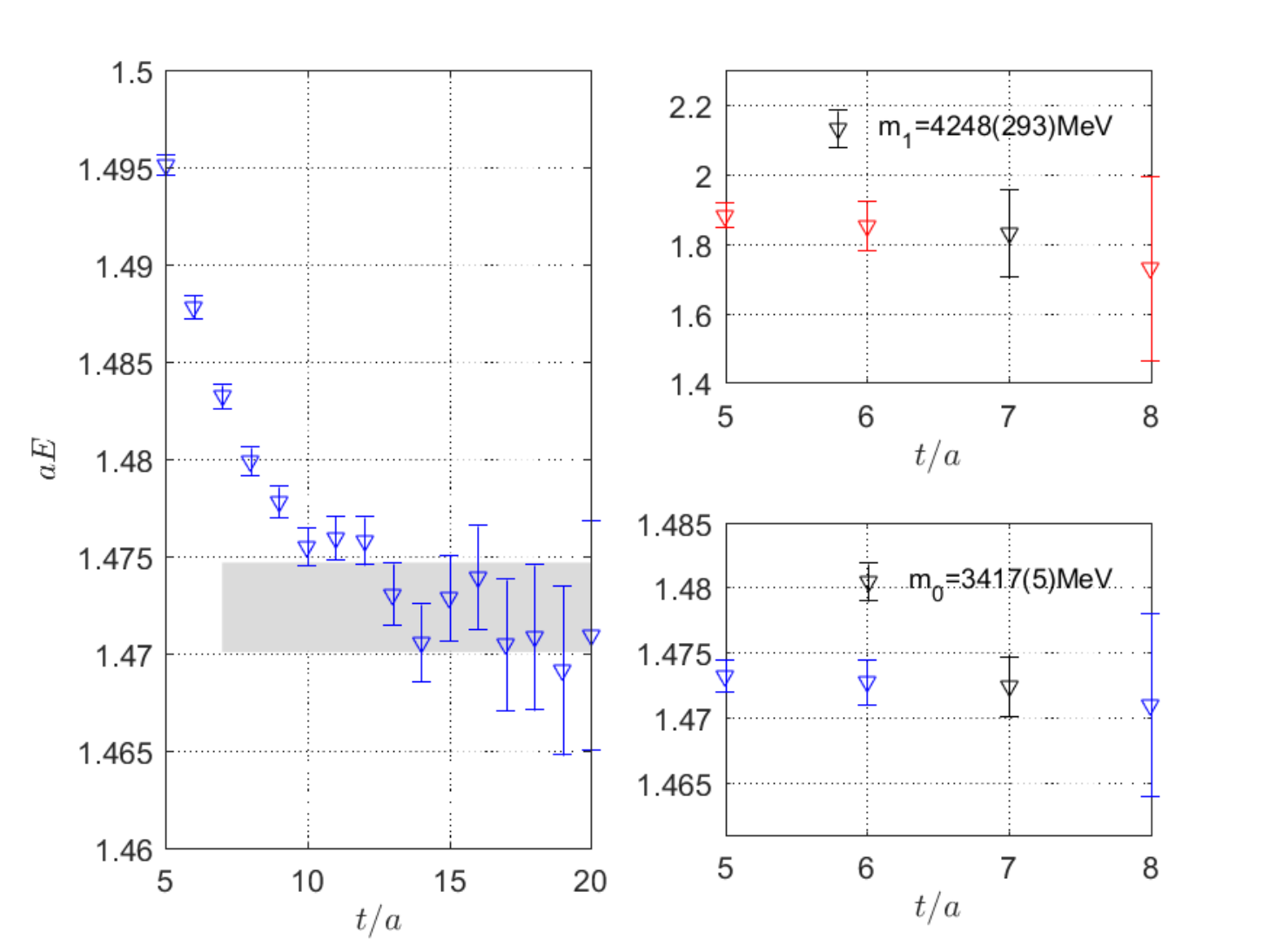}}
  \center (b) Ens.II
\end{minipage}\\
\begin{minipage}{0.95\linewidth}
  %\centerline
  {\includegraphics[width=0.95\linewidth]{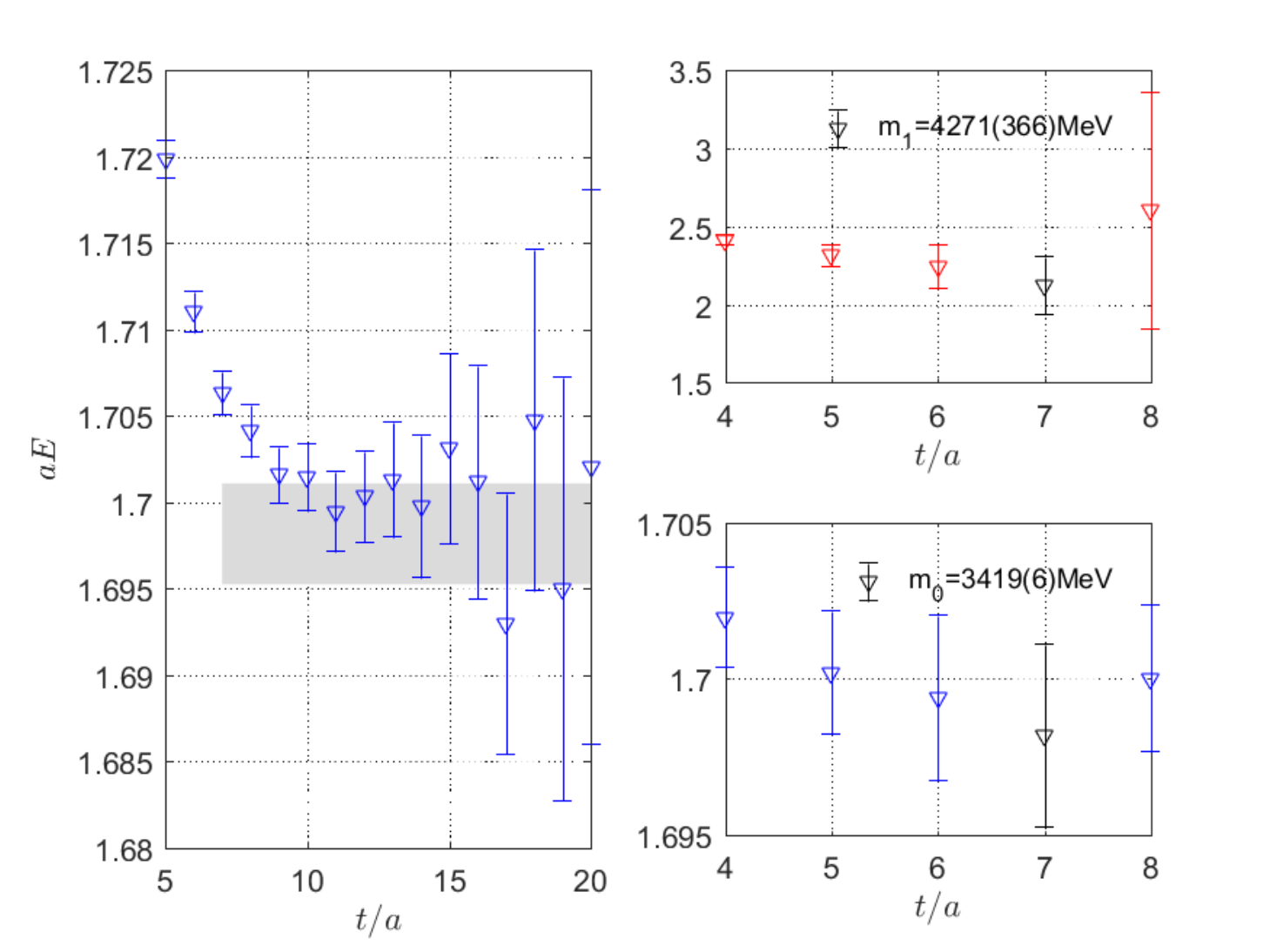}}
  \center (c) Ens.III
\end{minipage}
\caption{The left panels show the effective mass at different time slices together with the corresponding fitting ranges (grey bands) and the right panels are the ground and excited state mass values fitted from two-point correlation functions using Eq.~(\ref{eq:two-state-fit}).  The black symbols denote the chosen $m_0$, that correspond to the grey band to its left. The green symbols in (a) denote another choice for the $m_0$ and $m_1$.}
\label{fig:mass}
\end{figure}
 Since the generalized eigenvalues in Eq.~(\ref{eq:gevp}) decay exponentially,
 the corresponding mass eigenvalues can be extracted easily from,
\beq
\cosh(m_n)&=&\frac{\lambda_n(t-1)+\lambda_n(t+1)}{2\lambda_n(t)}.
\label{mass_ralation}
\eeq
 Since we want to extrapolate the form factor to eliminate the excited state contamination,
 we therefore use the following two-state fit form for the $\chi_{c0}$ correlator,
\be
\label{eq:two-state-fit}
C^{(2)}_{\chi_{c0}}(t)=V \sum_{i=0,1}\frac{Z_i^2}{2m_i} \left(e^{-m_it}+e^{-m_i(T-t)}\right)
\ee
with $V$ being the spatial volume, $m_0$ the ground state mass and $m_1$ the first excited state mass.
 The factors $Z_i=\frac{1}{\sqrt{V}}\langle i|\mathcal{O}_{\chi_{c0}}^\dagger|0\rangle$ (with $i=0,1$)
 are the overlap amplitudes for the ground and the first excited state, respectively.
 The corresponding mass plateaus and the  masses are illustrated in Fig.~\ref{fig:mass} for the
 three ensembles we utilize in this work. The left column of the panels show the effective mass on each time slice. The right panels denote the mass values fitted from two-point correlation functions, the upper one for the first excited state and the bottom one for the ground state.
 As the grey bands in the left panels indicate, the starting time slices are adjusted according to $\chi^2/d.o.f$ of the fit while the ending time slices are fixed to be $t_{\max} = 27, 20, 20$ for ensemble I, II, III, respectively.
 Noting that the grey band of Ens.I is obviously different from the other two ensembles, the ground state mass $m_0$ might be underestimated. So we calculate the result for another plateau with green mark and take the difference of them as the major source of systematic uncertainty.
% \textcolor{blue}{What are the data points in the left panels??Need to explain here.}

  The results for the mass values are summarized in Table~\ref{tab:mass}.
  Note that we use the $\eta_c$ mass to fix the valence charm quark mass $a\mu_c$ in this work. And the $\chi_{c0}$ experiment mass is $3414.7(3)$MeV
  quoted by PDG~\cite{Zyla:2020zbs}.
\begin{table}[!htb]
\begin{center}
\caption{Mass value $m_0$ and spectral weight $Z_0$ for ground state and the first excited state mass $m_1$ on each ensemble respectively. Ens.I(a) and Ens.I(b) are corresponding to the black and green symbols respectively.}
%The last line cites the corresponding result from PDG~\cite{pdg2021}.}
\begin{ruledtabular}
\begin{tabular}{cccc}
                           &$m_0$[MeV]    &$Z_0$     &$m_1$[MeV]         \\
\hline
Ens.I(a)                      &3438(9)        &0.0959(25)   &3906(176)           \\
\hline
Ens.I(b)                      &3445(4)        &0.0972(9)   &4181(57) \\
\hline
Ens.II                     &3417(5)        &0.1216(10)    &4248(293)          \\
\hline
Ens.III                    &3419(6)       &0.1320(7)  & 4271(366)          \\
\end{tabular}
\end{ruledtabular}
\label{tab:mass}
\end{center}
\end{table}

 \begin{figure*}[!htb]
\begin{minipage}{0.495\linewidth}
  %\centerline
  {\includegraphics[width=9cm]{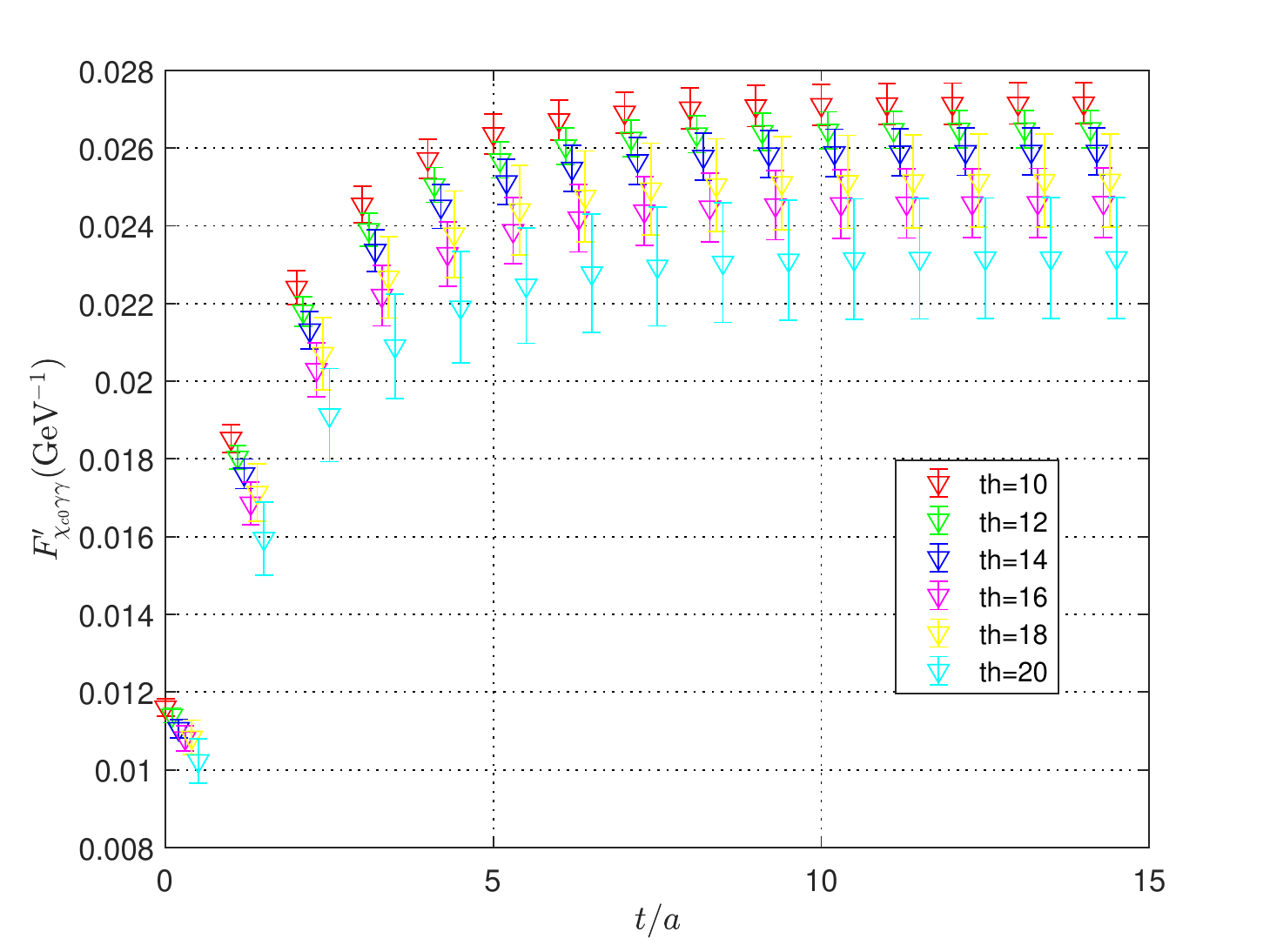}}
  %\centerline{$n_2=(0 -1 -2)$; $n_f=(0\ 0\ 0)$}
  %\center\ \ \ \ \ \ \ \ \ \ \ \ \ \ \
  \center (a)
\end{minipage}
\hfill
\begin{minipage}{0.495\linewidth}
  %\centerline
  {\includegraphics[width=9cm]{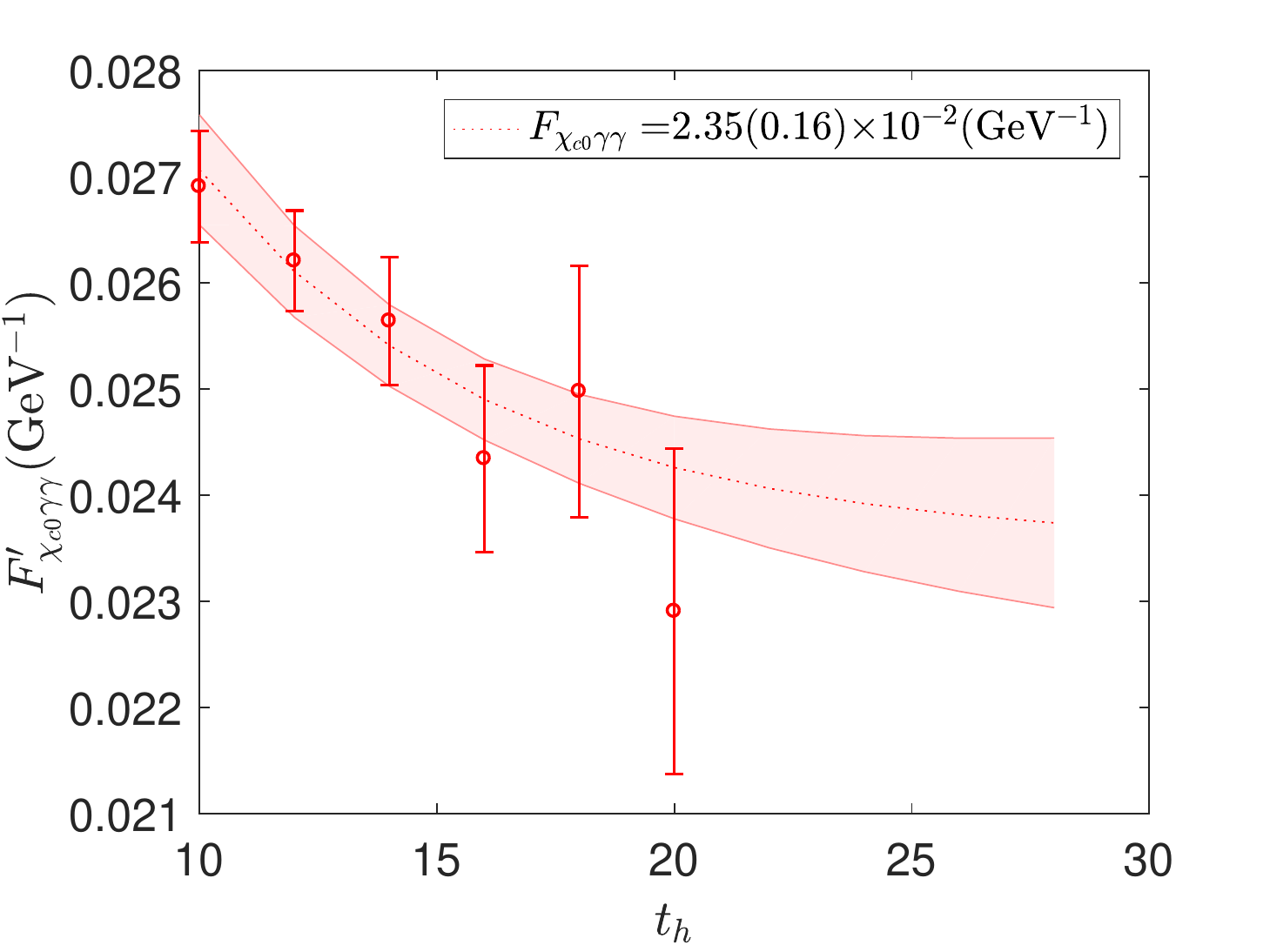}}
  %\centerline{$n_2=(0 -1 -2)$; $n_f=(0\ 0\ 0)$}
  %\center\ \ \ \ \ \ \ \ \ \ \ \ \ \ \
  \center (b)
\end{minipage}\\

\begin{minipage}{0.495\linewidth}
  %\centerline
  {\includegraphics[width=9cm]{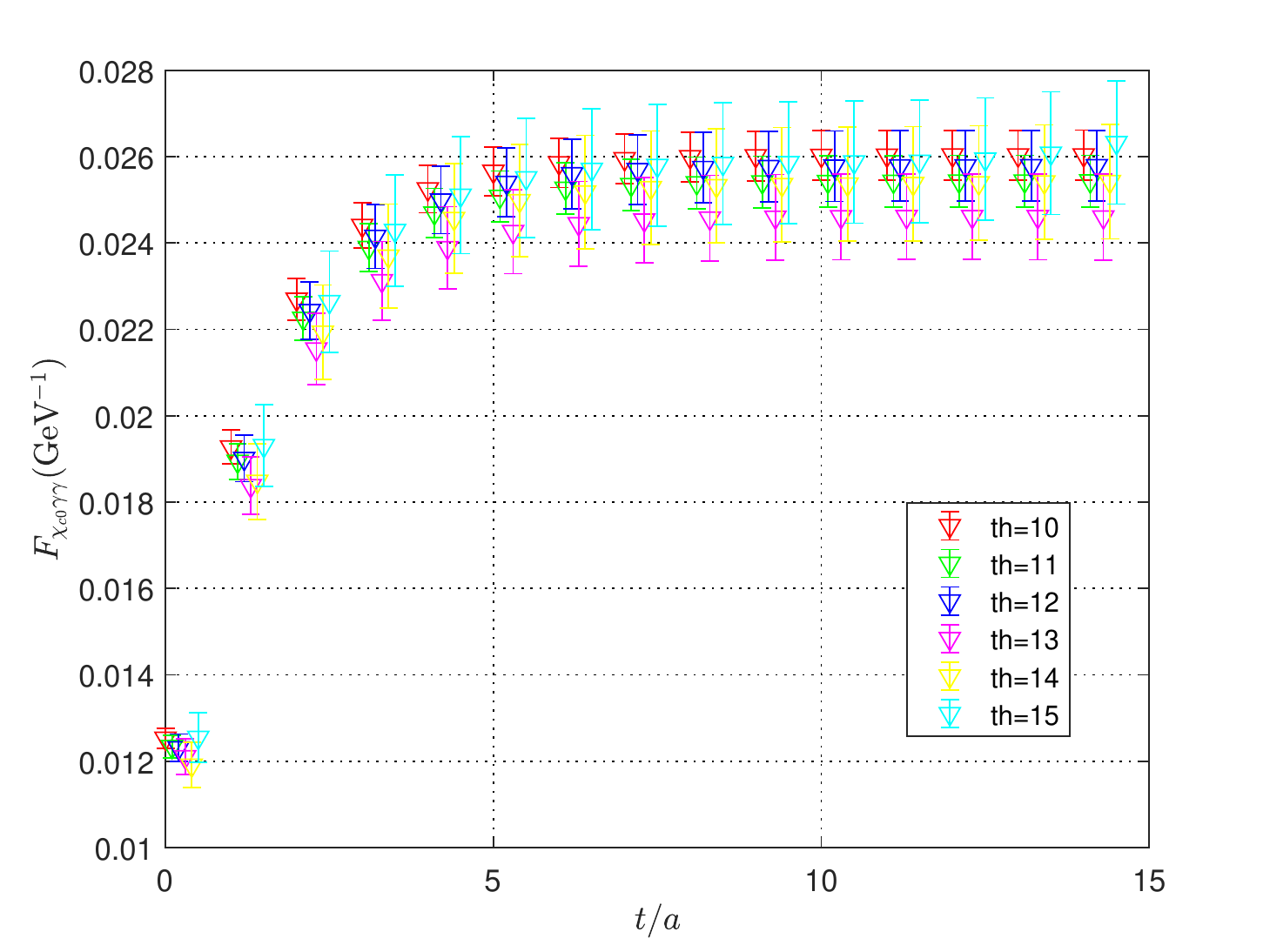}}
  %\centerline{$n_2=(0 -1 -2)$; $n_f=(0\ 0\ 0)$}
  %\center\ \ \ \ \ \ \ \ \ \ \ \ \ \ \
  \center (c)
\end{minipage}
\hfill
\begin{minipage}{0.495\linewidth}
  %\centerline
  {\includegraphics[width=9cm]{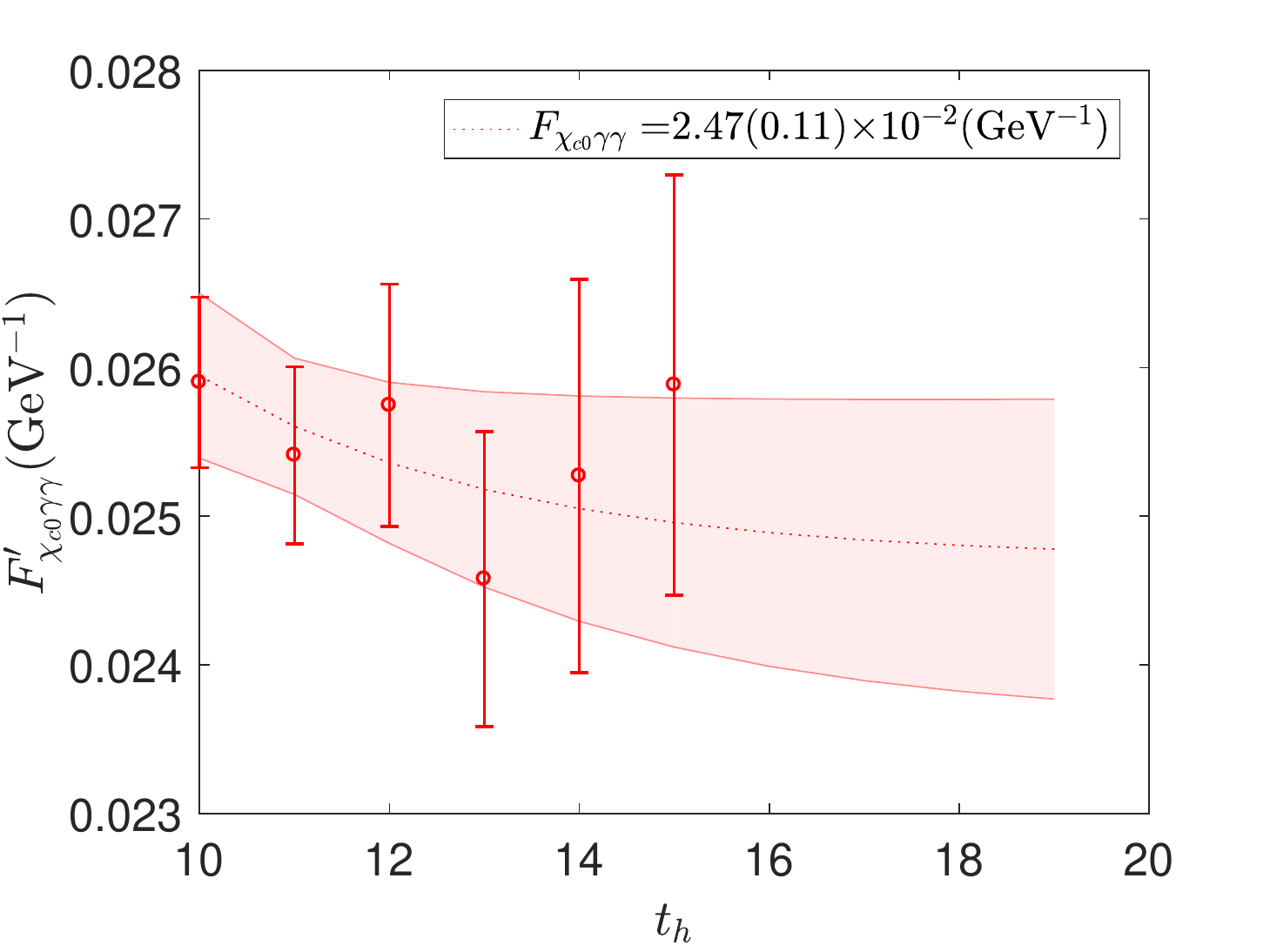}}
  %\centerline{$n_2=(0 -1 -2)$; $n_f=(0\ 0\ 0)$}
  %\center\ \ \ \ \ \ \ \ \ \ \ \ \ \ \
  \center (d)
\end{minipage}\\

\begin{minipage}{0.495\linewidth}
  %\centerline
  {\includegraphics[width=9cm]{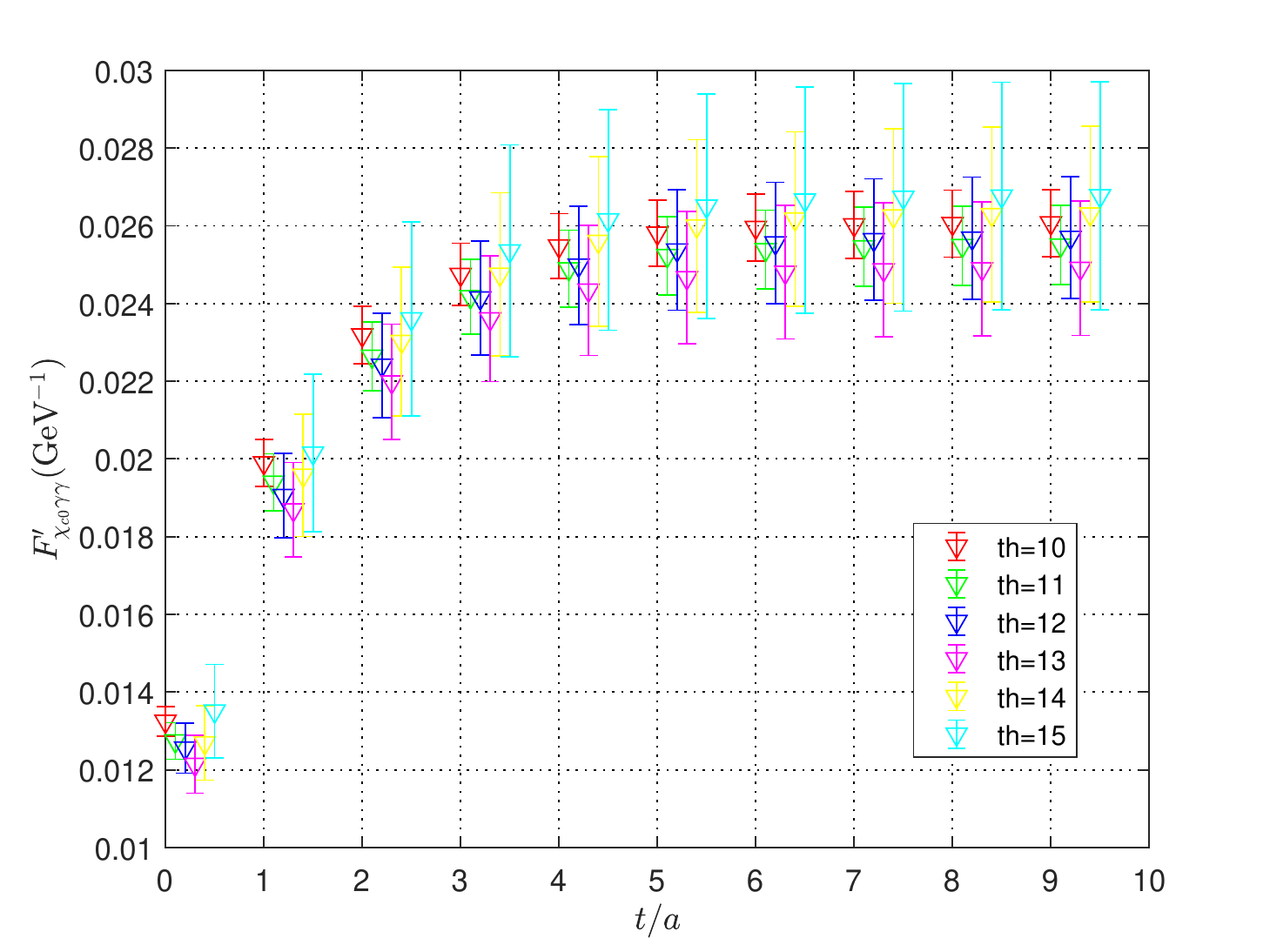}}
  %\centerline{$n_2=(0 -1 -2)$; $n_f=(0\ 0\ 0)$}
  %\center\ \ \ \ \ \ \ \ \ \ \ \ \ \ \
  \center (e)
\end{minipage}
\hfill
\begin{minipage}{0.495\linewidth}
  %\centerline
  {\includegraphics[width=9cm]{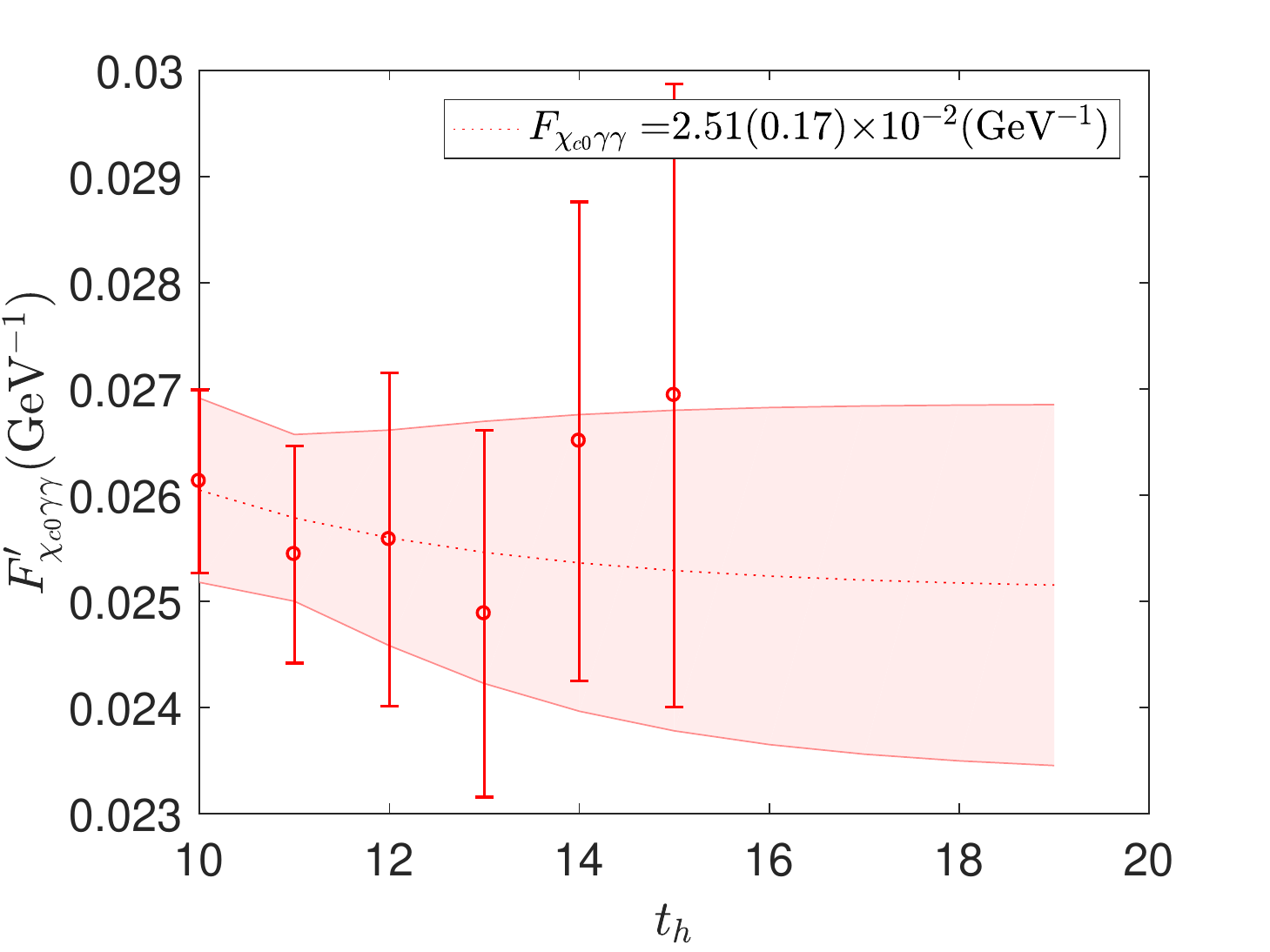}}
  %\centerline{$n_2=(0 -1 -2)$; $n_f=(0\ 0\ 0)$}
  %\center\ \ \ \ \ \ \ \ \ \ \ \ \ \ \
  \center (f)
\end{minipage}\\
\caption{The left column represents the plateaus of the form factor with different $t_h$ while the right column shows the extrapolation to the ground state contribution. The label (a) and (b), (c) and (d), (e) and (f) are for Ens.I(a), Ens.II, Ens.III respectively.}
\label{fig:form}
\end{figure*}

\subsection{Renormalization factor $Z_V$}
\label{sec:Z}
The hadronic tensor $\mathcal{H}_{\mu\nu}$ contains the electromagnetic current operators $J_\mu$ from all flavor of quarks.
However, since we neglect the disconnected diagrams in this study, we only need to consider the charm quark current $J^{(c)}_\mu=\bar{c}\gamma_\mu c(\vec{x},t)$.
Since we adopt the local current form, there exists an extra multiplicative renormalization factor $Z_V$
that can be calculated by a ratio of the two-point function and the three-point function as in Eq.~(\ref{eq:ZV}).
In principle this renormalization factor does not depend on the particle state used to calculate it.
 For a better signal, we choose to use the $\eta_c$ correlators instead of $\chi_{c0}$.
Taking account of the around-of-world effect,
we use the following relation to extract $Z_V$.
\be
\label{eq:ZV}
Z_V=\frac{\sum_{\vec{x}}\langle \mathcal{O}_{\eta_c}(t)\mathcal{O}^\dagger_{\eta_c}(0)\rangle}{\sum_{\vec{x}}\langle \mathcal{O}_{\eta_c}(t)J_0^{(c)}(t/2,\vec{x})\mathcal{O}^\dagger_{\eta_c}(0)\rangle}\frac{1}{(1+e^{-m_0(T-2t)})}
\ee
The results for $Z_V$ are listed in Table~\ref{tab:ZV}.
\begin{table}[!htb]
\begin{tabular}{cccc}
\hline
\hline
 & Ens.I & Ens.II & Ens.III   \\
\hline
$Z_V$  &  0.6523(21)  & 0.6296(29) &  0.6057(27)\\
\hline
\hline
\end{tabular}
\caption{Renormalization factor $Z_V$ for three ensembles.}
\label{tab:ZV}
\end{table}

 When computing the scalar form factor $F_{\chi_{c0}\gamma\gamma}$ in Eq.~(\ref{eq:Fchi}) on the lattice,
 the integration over space-time are replaced by discrete summations.
 When two identical currents in $\mathcal{H}_{\mu\nu}(t,\vec{x})$, meaning that they share the same Lorentz index, are at the same space-time point,
 an extra renormalization is needed to take into account of the contact term. This is due to a new type of composite operator that
 is not properly renormalized yet, even if each current is already properly renormalized by the factor $Z_V$.
 In order to take this effect into account, one needs to impose another appropriate renormalization condition for this new composite operator.
 In this work, we choose not to sum the same space-time point contributions for identical currents
 and thereby avoiding this potential renormalization.
 To summarize, the above mentioned procedures taken on already $O(a)$-improved
ensembles will at most introduce an extra $O(a^2)$ lattice artifact on
physical observables which will be taken care of in the final continuum extrapolation.

\subsection{The scalar form factor $F_{\chi_{c0}\gamma\gamma}$}
\label{sec:form factor}

 When computing the hadronic tensor, we evaluate the three-point correlation function $\langle J_\mu(x)J_\nu(0)\mathcal{O}^\dagger_{\chi_{c0}}(-t_h)\rangle$.
 To produce the static meson state, we use the $Z_4$-stochastic wall source placed at time-slice $-t_h$.
 This cuts the uncertainty by nearly a half when compared with the simple point source for the meson mass.
 We also apply the APE~\cite{ape} and Gaussian smearing~\cite{Gusken:1989qx} for the gauge field and $\chi_{c0}$ operator.
 We utilize the random point source propagator for the current to
 arrive at the three-point correlation function.
 In practice, the hadronic tensor with current $J_\nu(0)$ placed at zero point is
 actually an average of all the time slices and a random positions on each time slice.

 Consequently, the scalar form factor we computed according to Eq.~(\ref{eq:Fchi}) on the lattice $F'_{\chi_{c0}\gamma\gamma}$ actually suffers from excited state contamination due to higher excitation states of $\chi_{c0}$.
 What we really need is the ground state $\chi_{c0}$. This effect can be taken care of by
 considering $t_h$ dependence of the form factors. Therefore, we computed several different
 separations $t_h$ and perform the following fit,
\be
F'_{\chi_{c0}\gamma\gamma}(t_h)=\ F_{\chi_{c0}\gamma\gamma}+\xi\cdot e^{-(m_1-m_0)t_h}\;,
\ee
 where $F_{\chi_{c0}\gamma\gamma}$ and $\xi$ are the two free parameters.
 For the parameters $m_0$ and $m_1$, we take the values presented in Table~\ref{tab:mass}.
 %\textcolor{red}{So, what really goes into the Eq.(\ref{eq:gamma}) is $F_0$, and not $F_{\chi_{c0}\gamma\gamma}$?? If so, we'd better use another symbol for these.... }
 The form factor with different time separation $t_h$ together with the ground state extrapolation values for $F_{\chi_{c0}\gamma\gamma}$
 for three set of ensembles are illustrated  in Fig.~\ref{fig:form}.

\subsection{Comparison of the form factor with previous lattice results}
\label{subsec:comparison_with_latt}

%  \begin{widetext}
\begin{table*}[!htb]
%\scriptsize
\begin{tabular}{c|cc|c|c}
\hline\hline
$G(0,0)$ & Ens.I(a) &Ens.I(b) & Ens.II & Ens.III   \\
\hline
  This work  &  0.1884(123) &  0.1899(69) & 0.1911(85) & 0.1931(131) \\
 \hline
Ref.\cite{Chen_2020} & \multicolumn{2}{c|}{0.09079(8)(19)(90)}  & 0.1017(7)(102)(126)& -\\
 \hline\hline
\end{tabular}
\caption{Dimensionless form factors $G(0,0)$ obtained in this work and those
 obtained in Ref.~\cite{Chen_2020} for each ensemble.  Ens.I(a) and Ens.I(b) denotes two
 different results obtained by taking two different $\chi_{c0}$ mass
 values as discussed in Fig.~\ref{fig:mass}. Ensemble III was not available in Ref.\cite{Chen_2020}.
 Errors quoted for $G(0,0)$ in this work are purely statistical that are obtained using 
 the conventional jackknife method.
 Three errors for the results from Ref.\cite{Chen_2020} stands for errors from statistical,
 from momentum extrapolations and estimates for the finite lattice spacings.
}
\label{tab:width}
\end{table*}
%\end{widetext}

  The most recent lattice computation on $\chi_{c0}\rightarrow\gamma\gamma$ decay in the literature
  is the one from CLQCD~\cite{Chen_2020}, which happened to use exactly the same set of ensembles
  as this work. This allows a more detailed comparison on the level of dimensionless form factors
  for each of the common lattice spacings. For this purpose, we decide to convert our results
  for $F_{\chi_{c0}\gamma\gamma}$ into dimensionless quantities which could be taken as
  either $\Gamma_{\gamma\gamma}(\chi_{c0})/m_{\chi_{c0}}$ or the dimensionless form factor $G(0,0)$ that
  is utilized in Ref.~\cite{Chen_2020}.  The relation between these two dimensionless quantities 
  is easily found to be,
  \begin{equation}
  \label{eq:dimensionless-ratios}
  \!\!\! \frac{\Gamma_{\gamma\gamma}(\chi_{c0})}{m_{\chi_{c0}}}
  =\alpha^2 \pi |m_{\chi_{c0}}F_{\chi_{c0}\gamma\gamma}|^2
  =\alpha^2 \pi (e_c)^4|G(0,0)|^2.
  \end{equation}
  Dimensionless quantities have the advantage that they are independent of the scale setting
  process for the lattice spacings, which is subject to its own errors depending on how the scale was set.
  Since the scale setting processes for lattice calculations also progress over the years, 
  the information for the lattice spacing in physical units, 
  both the central values and the errors,  are also changing with time even for a given particular ensemble. 
  It is therefore better to attach these errors due to scale-setting at the
  very end when comparing with the experiments. In the intermediate step when comparing
  with other lattice computations, it is easier to directly compare the dimensionless quantities if possible. 
  In fact, this allows us to compare with previous lattice results in Ref.~\cite{Chen_2020} 
  at each individual lattice spacing, namely Ens.I and Ens.II that have also been utilized. 
  Of course, when quoting the final physical decay width,
  the effect of scale setting will be taken into account together with its associated errors.

 In Table~\ref{tab:width}, the dimensionless form factor $G(0,0)$ obtained 
 via Eq.~(\ref{eq:dimensionless-ratios}) from $F_{\chi_{c0}\gamma\gamma}$  
 for all three ensembles are listed together with the corresponding results for
  Ens.I and Ens.II from Ref.~\cite{Chen_2020}. 
  Ens.III was not utilized in the study of Ref.~~\cite{Chen_2020}.
  Two entries for Ens.I, labelled as Ens.I(a) and Ens.I(b)
  corresponds to the two different ways of extracting $\chi_{c0}$ masses as discusses in Fig.~\ref{fig:mass}.
  The errors quoted for $G(0,0)$ in this work are obtained using the conventional jackknife method.
  As for the three errors for the results from Ref.~\cite{Chen_2020}, 
  they stand for errors from statistical, from momentum extrapolations and 
  estimates of the finite lattice spacing errors, respectively. 
  We notice that the central values for dimensionless form factors $G(0,0)$ 
  differ by almost a factor of two for Ens.I and Ens.II.
  The reason of this apparent discrepancy is still unknown to us. 
  One possibility could be the under estimation of the lattice artifacts 
  for each of the ensemble in Ref.~\cite{Chen_2020}.

\subsection{Continuum extrapolation and the final result and discussions}
\label{sec:final result}

\begin{figure}[!h]
	\centering
		\subfigure{\includegraphics[width=0.48\textwidth]{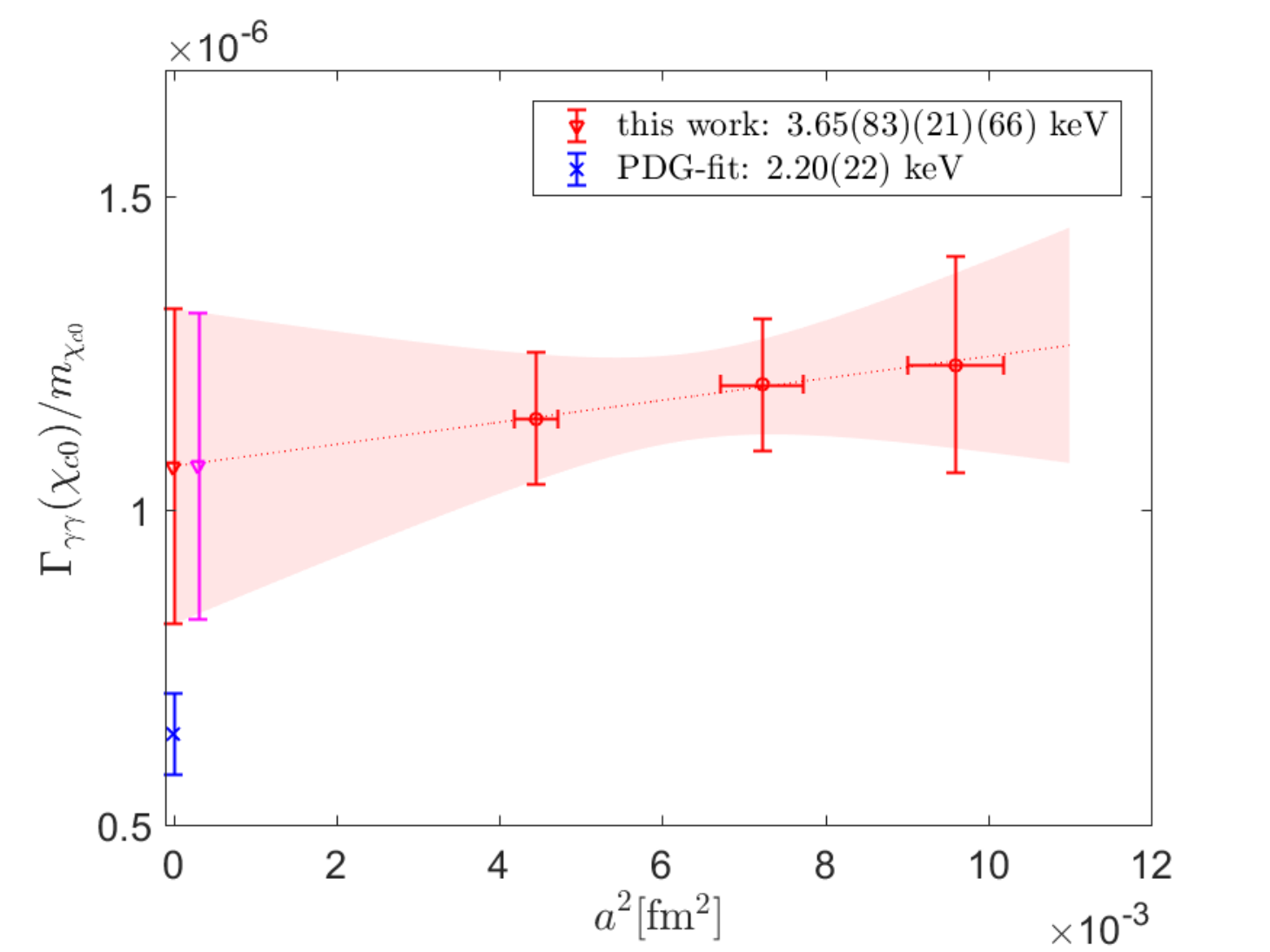}}
\caption{Continuum extrapolation for the ratio $\Gamma_{\gamma\gamma}(\chi_{c0})/m_{\chi_{c0}}$. 
 The three data points with both horizontal and vertical error-bars are results
 from three ensembles. 
 The extrapolated results are shown by two side-by-side points near $a^2=0$: 
 The one with a smaller error bar (the right one) represents
  the extrapolation result without considering lattice spacing errors. 
  The other one (left one) is the result with lattice spacing errors taken into consideration. 
  The data point (blue) below these two with a smaller error is  the PDG-fit value
  for this ratio. At the upper right corner, we have also indicated the result
  of the width in physical units.}
\label{fig:explt}
\end{figure}

 After obtaining the dimensionless form factors for three different lattice spacings, 
 we could investigate the continuum limit of this quantity. For this purpose, 
 we decide to perform this extrapolation using the more physical quantity $\Gamma_{\gamma\gamma}(\chi_{c0})/m_{\chi_{c0}}$, 
 which is proportional to the norm-squared dimensionless form factor $|G(0,0)|^2$  as 
 indicated in Eq.~(\ref{eq:dimensionless-ratios}).
 The continuum extrapolation is done by performing a linear fit in $a^2$ for the three ensembles
 and the result after the continuum extrapolation, together with the results for each ensemble,
 are illustrated in Fig.~\ref{fig:explt}.
 Here the horizontal error bars for the data points indicate the errors in $a^2$ inferred from Ref.~\cite{BOUCAUD2008695,tw2010}
 
 It is seen that the three data points fit nicely on a straight line yielding a
 reasonable $\chi^2/d.o.f$. The two points near $a^2=0$ with larger error bars designate
 two different results obtained from the fit with and without considering 
 the horizontal $a^2$-errors for the lattice spacings. Below the two data points,
 we also plot the corresponding experimental value from PDG for this ratio. 
 The two extrapolated results share almost identical central values. 
 They differ only by their errors. The point with slightly
 larger error (the one slightly to the left) is the one that takes into
 account of the horizontal $a^2$-errors while the other one is the one without considering $a^2$-errors. 
 Finally, there is another source of systematic errors arising from the different plateaus 
 in the mass as discussed in Sec.~\ref{sec:mass}. Therefore, 
 we finally quote the result of the decay width in physical units as
\begin{equation}
\label{eq:finalresult}
\Gamma_{\gamma\gamma}(\chi_{c0})=3.65(83)_{\mathrm{stat}}(21)_{\mathrm{lat.syst}}(66)_{\mathrm{syst}}\, \textrm{keV},
\end{equation}
 where the first two errors represent the error obtained without/with the $a^2$-errors.
 It should be interpreted as follows: the first error is the error without considering $a^2$-errors. 
 The second one with the subscript $\mathrm{lat.syst}$ indicate the extra amount of error 
 if one would consider the $a^2$-errors. In other words, one could add the first two errors in quadrature to obtain
 the error with $a^2$-errors taken into consideration, which is shown by
 the left point near $a^2=0$ in Fig.~\ref{fig:explt}.
 The last error with subscript $\mathrm{syst}$ reflects the systematic error 
 from different mass plateaus in Ens.I(a) and Ens.I(b).
 In this manner, we separate different sources of systematic errors that have been
 studied in this paper.
 
 It is evident that the central value for the decay width obtained in this paper 
 is larger than the PDG value.  But due to our large statistical and systematical uncertainty,
 it is still compatible with the experimental results within 1.3$\sigma$.

 We have tried to estimate the systematic uncertainties that might influence 
 our final result quoted above in Eq.~(\ref{eq:finalresult}).
 This includes choosing different plateaus for the mass, for the renormalization factor $Z_V$,
 spectral weight factor $Z_i$, and the number of time-slices we choose for the extrapolation 
 of the ground state form factor, etc.
 It turns out that only the two plateaus presented in Sec.~\ref{sec:mass} contribute to a visible 
 deviation in the central value, which we add in the third error in Eq.~\ref{eq:finalresult}.

 Needless to say, there are also other source of systematic errors that are more difficult to quantify,
 say neglecting the disconnected contributions, quenching of the the strange and charm quarks, etc.
 The disconnected diagrams contributions are believed to be suppressed in the charmonium system~\cite{ukqcd,taro}
 due to the Okubo-Zweig-Iizuka(OZI) rule.
 Furthermore, the non-physical masses of up and down quarks usually only result in a small effect which is indicated
 in the previous lattice calculations~\cite{bali}. 
 Therefore, the major direction in future improvements points to the
 deduction of the statistical noise in $\chi_{c0}$ correlation functions.
 Only after the large statistical uncertainty is fully under control,
 should we worry about other remaining systematic effects.

 Part of the large statistical error in our study can be traced back to the mixing of $\chi_{c0}$ and $\eta_c$
 in the twisted-mass formulation of lattice QCD.
 To entangle this mixing, we have utilized a GEVP procedure that projects out the operators best overlapped with $\eta_c$ and $\chi_{c0}$ as discussed in Sec.~\ref{sec:lattice setup}.
 Although this procedure works perfectly for the ground state $\eta_c$,
 the efficiency for $\chi_{c0}$ is not quite satisfactory,
 rendering the two-point and three-point correlation functions of $\chi_{c0}$ much noisier
 than that of $\eta_c$, resulting in a much larger error for the decay rate of $\chi_{c0}$.
 Possibilities to get around this difficulty could be simply increasing the statistics of the ensembles,
 using more interpolating operators as the basis operators
 or simply using a formulation that does not suffer from this mixing effect at all,
 e.g. utilizing the clover-improved Wilson fermion configurations.

\section{Conclusion}
 In this paper,  we report a new lattice QCD computation of the 
 scalar charmonium $\chi_{c0}$ to two-photon decay width.
 We have performed our study using three ensembles of $N_f=2$ twisted mass gauge field configurations at three
 different lattice spacings. This allows us to perform a more reliable continuum extrapolation therefore
 eliminating the substantial finite lattice spacing errors observed in previous lattice studies.
 We also adopt a new method that directly extracts the relevant on-shell form factor, by-passing
  the extrapolation in the photon virtualities.  We obtain the decay width of $\chi_{c0}$ meson
  to be $\Gamma_{\gamma\gamma}(\chi_{c0})=3.65(83)_{\mathrm{stat}}(21)_{\mathrm{lat.syst}}(66)_{\mathrm{syst}}\, \textrm{keV}$.
  Albeit the large errors in this computation, the result is  
  compatible with the existing experimental values within 1.3$\sigma$.
 Further possible improvements are also discussed.  This calculation and possible future more systematic studies will
 await the new experimental results that will become available soon.

\section*{Acknowledgments}
The authors would like to thank the Extended Twisted Mass Collaboration(ETMC)
for sharing the gauge configurations with us.
C.L. and Z.H.Z. acknowledge the support by NSFC of China under Grant No. 12070131001 and also the  support by CAS Interdisciplinary Innovation Team and NSFC of China under Grant No. 11935017.
Y.M. are supported by NSFC of China under Grant No. 12047505 and State Key Laboratory of Nuclear Physics and Technology, Peking University.
The calculations were carried out on High-performance Computing Platform of Peking University and Tianhe-1A supercomputer at Tianjin National Supercomputing Center.

%\bibliography{ref}

\begin{thebibliography}{36}%
\makeatletter
\providecommand \@ifxundefined [1]{%
 \@ifx{#1\undefined}
}%
\providecommand \@ifnum [1]{%
 \ifnum #1\expandafter \@firstoftwo
 \else \expandafter \@secondoftwo
 \fi
}%
\providecommand \@ifx [1]{%
 \ifx #1\expandafter \@firstoftwo
 \else \expandafter \@secondoftwo
 \fi
}%
\providecommand \natexlab [1]{#1}%
\providecommand \enquote  [1]{``#1''}%
\providecommand \bibnamefont  [1]{#1}%
\providecommand \bibfnamefont [1]{#1}%
\providecommand \citenamefont [1]{#1}%
\providecommand \href@noop [0]{\@secondoftwo}%
\providecommand \href [0]{\begingroup \@sanitize@url \@href}%
\providecommand \@href[1]{\@@startlink{#1}\@@href}%
\providecommand \@@href[1]{\endgroup#1\@@endlink}%
\providecommand \@sanitize@url [0]{\catcode `\\12\catcode `\$12\catcode
  `\&12\catcode `\#12\catcode `\^12\catcode `\_12\catcode `\%12\relax}%
\providecommand \@@startlink[1]{}%
\providecommand \@@endlink[0]{}%
\providecommand \url  [0]{\begingroup\@sanitize@url \@url }%
\providecommand \@url [1]{\endgroup\@href {#1}{\urlprefix }}%
\providecommand \urlprefix  [0]{URL }%
\providecommand \Eprint [0]{\href }%
\providecommand \doibase [0]{https://doi.org/}%
\providecommand \selectlanguage [0]{\@gobble}%
\providecommand \bibinfo  [0]{\@secondoftwo}%
\providecommand \bibfield  [0]{\@secondoftwo}%
\providecommand \translation [1]{[#1]}%
\providecommand \BibitemOpen [0]{}%
\providecommand \bibitemStop [0]{}%
\providecommand \bibitemNoStop [0]{.\EOS\space}%
\providecommand \EOS [0]{\spacefactor3000\relax}%
\providecommand \BibitemShut  [1]{\csname bibitem#1\endcsname}%
\let\auto@bib@innerbib\@empty
%</preamble>
\bibitem [{\citenamefont {Huang}\ \emph {et~al.}(1996)\citenamefont {Huang},
  \citenamefont {Qiao},\ and\ \citenamefont {Chao}}]{huang96}%
  \BibitemOpen
  \bibfield  {author} {\bibinfo {author} {\bibfnamefont {H.-W.}\ \bibnamefont
  {Huang}}, \bibinfo {author} {\bibfnamefont {C.-F.}\ \bibnamefont {Qiao}},\
  and\ \bibinfo {author} {\bibfnamefont {K.-T.}\ \bibnamefont {Chao}},\
  }\bibfield  {title} {\bibinfo {title} {Electromagnetic annihilation rates of
  ${\ensuremath{\chi}}_{c0}$ and ${\ensuremath{\chi}}_{c2}$ with both
  relativistic and qcd radiative corrections},\ }\href
  {https://doi.org/10.1103/PhysRevD.54.2123} {\bibfield  {journal} {\bibinfo
  {journal} {Phys. Rev. D}\ }\textbf {\bibinfo {volume} {54}},\ \bibinfo
  {pages} {2123} (\bibinfo {year} {1996})}\BibitemShut {NoStop}%
\bibitem [{\citenamefont {Hwang}\ and\ \citenamefont {Guo}(2010)}]{hwang10}%
  \BibitemOpen
  \bibfield  {author} {\bibinfo {author} {\bibfnamefont {C.-W.}\ \bibnamefont
  {Hwang}}\ and\ \bibinfo {author} {\bibfnamefont {R.-S.}\ \bibnamefont
  {Guo}},\ }\bibfield  {title} {\bibinfo {title} {Two-photon and two-gluon
  decays of $p$-wave heavy quarkonium using a covariant light-front approach},\
  }\href {https://doi.org/10.1103/PhysRevD.82.034021} {\bibfield  {journal}
  {\bibinfo  {journal} {Phys. Rev. D}\ }\textbf {\bibinfo {volume} {82}},\
  \bibinfo {pages} {034021} (\bibinfo {year} {2010})}\BibitemShut {NoStop}%
\bibitem [{\citenamefont {Ecklund}\ \emph {et~al.}(2008)\citenamefont {Ecklund}
  \emph {et~al.}}]{cleo}%
  \BibitemOpen
  \bibfield  {author} {\bibinfo {author} {\bibfnamefont {K.~M.}\ \bibnamefont
  {Ecklund}} \emph {et~al.} (\bibinfo {collaboration} {CLEO Collaboration}),\
  }\bibfield  {title} {\bibinfo {title} {Two-photon widths of the
  ${\ensuremath{\chi}}_{cJ}$ states of charmonium},\ }\href
  {https://doi.org/10.1103/PhysRevD.78.091501} {\bibfield  {journal} {\bibinfo
  {journal} {Phys. Rev. D}\ }\textbf {\bibinfo {volume} {78}},\ \bibinfo
  {pages} {091501} (\bibinfo {year} {2008})}\BibitemShut {NoStop}%
\bibitem [{\citenamefont {Ablikim}\ \emph {et~al.}(2017)\citenamefont {Ablikim}
  \emph {et~al.}}]{bes}%
  \BibitemOpen
  \bibfield  {author} {\bibinfo {author} {\bibfnamefont {M.}~\bibnamefont
  {Ablikim}} \emph {et~al.} (\bibinfo {collaboration} {BESIII Collaboration}),\
  }\bibfield  {title} {\bibinfo {title} {Improved measurements of two-photon
  widths of the ${\ensuremath{\chi}}_{cJ}$ states and helicity analysis for
  ${\ensuremath{\chi}}_{c2}\ensuremath{\rightarrow}\ensuremath{\gamma}\ensuremath{\gamma}$},\
  }\href {https://doi.org/10.1103/PhysRevD.96.092007} {\bibfield  {journal}
  {\bibinfo  {journal} {Phys. Rev. D}\ }\textbf {\bibinfo {volume} {96}},\
  \bibinfo {pages} {092007} (\bibinfo {year} {2017})}\BibitemShut {NoStop}%
\bibitem [{\citenamefont {Appelquist}\ and\ \citenamefont
  {Politzer}(1975)}]{appel}%
  \BibitemOpen
  \bibfield  {author} {\bibinfo {author} {\bibfnamefont {T.}~\bibnamefont
  {Appelquist}}\ and\ \bibinfo {author} {\bibfnamefont {H.~D.}\ \bibnamefont
  {Politzer}},\ }\bibfield  {title} {\bibinfo {title} {Heavy quarks and
  ${e}^{+}{e}^{\ensuremath{-}}$ annihilation},\ }\href
  {https://doi.org/10.1103/PhysRevLett.34.43} {\bibfield  {journal} {\bibinfo
  {journal} {Phys. Rev. Lett.}\ }\textbf {\bibinfo {volume} {34}},\ \bibinfo
  {pages} {43} (\bibinfo {year} {1975})}\BibitemShut {NoStop}%
\bibitem [{\citenamefont {Barnes}()}]{Barnes}%
  \BibitemOpen
  \bibfield  {author} {\bibinfo {author} {\bibfnamefont {T.}~\bibnamefont
  {Barnes}},\ }\href@noop {} {\bibinfo  {journal} {{\em Proceedings of the IX
  International Workshop on Photon-Photon Collisisons}, edited by D. O.
  Caldwell and H. P. Paar(World Scientific, Singapore, 1992) p. 263}\
  }\BibitemShut {NoStop}%
\bibitem [{\citenamefont {Gupta}\ \emph {et~al.}(1996)\citenamefont {Gupta},
  \citenamefont {Johnson},\ and\ \citenamefont {Repko}}]{Gupta}%
  \BibitemOpen
\bibfield  {journal} {  }\bibfield  {author} {\bibinfo {author} {\bibfnamefont
  {S.~N.}\ \bibnamefont {Gupta}}, \bibinfo {author} {\bibfnamefont {J.~M.}\
  \bibnamefont {Johnson}},\ and\ \bibinfo {author} {\bibfnamefont {W.~W.}\
  \bibnamefont {Repko}},\ }\bibfield  {title} {\bibinfo {title} {Relativistic
  two-photon and two-gluon decay rates of heavy quarkonia},\ }\href
  {https://doi.org/10.1103/PhysRevD.54.2075} {\bibfield  {journal} {\bibinfo
  {journal} {Phys. Rev. D}\ }\textbf {\bibinfo {volume} {54}},\ \bibinfo
  {pages} {2075} (\bibinfo {year} {1996})}\BibitemShut {NoStop}%
\bibitem [{\citenamefont {Ebert}\ \emph {et~al.}(2003)\citenamefont {Ebert},
  \citenamefont {Faustov},\ and\ \citenamefont {Galkin}}]{ebert}%
  \BibitemOpen
  \bibfield  {author} {\bibinfo {author} {\bibfnamefont {D.}~\bibnamefont
  {Ebert}}, \bibinfo {author} {\bibfnamefont {R.}~\bibnamefont {Faustov}},\
  and\ \bibinfo {author} {\bibfnamefont {V.}~\bibnamefont {Galkin}},\
  }\bibfield  {title} {\bibinfo {title} {Two-photon decay rates of heavy
  quarkonia in the relativistic quark model},\ }\href
  {https://doi.org/10.1142/S021773230300971X} {\bibfield  {journal} {\bibinfo
  {journal} {Mod.Phys.Lett.A}\ }\textbf {\bibinfo {volume} {18}},\ \bibinfo
  {pages} {601} (\bibinfo {year} {2003})}\BibitemShut {NoStop}%
\bibitem [{\citenamefont {Godfrey}\ and\ \citenamefont
  {Isgur}(1985)}]{godfrey}%
  \BibitemOpen
  \bibfield  {author} {\bibinfo {author} {\bibfnamefont {S.}~\bibnamefont
  {Godfrey}}\ and\ \bibinfo {author} {\bibfnamefont {N.}~\bibnamefont
  {Isgur}},\ }\bibfield  {title} {\bibinfo {title} {Mesons in a relativized
  quark model with chromodynamics},\ }\href
  {https://doi.org/10.1103/PhysRevD.32.189} {\bibfield  {journal} {\bibinfo
  {journal} {Phys. Rev. D}\ }\textbf {\bibinfo {volume} {32}},\ \bibinfo
  {pages} {189} (\bibinfo {year} {1985})}\BibitemShut {NoStop}%
\bibitem [{\citenamefont {Bodwin}\ \emph {et~al.}(1992)\citenamefont {Bodwin},
  \citenamefont {Braaten},\ and\ \citenamefont {Lepage}}]{Bodwin}%
  \BibitemOpen
  \bibfield  {author} {\bibinfo {author} {\bibfnamefont {G.~T.}\ \bibnamefont
  {Bodwin}}, \bibinfo {author} {\bibfnamefont {E.}~\bibnamefont {Braaten}},\
  and\ \bibinfo {author} {\bibfnamefont {G.~P.}\ \bibnamefont {Lepage}},\
  }\bibfield  {title} {\bibinfo {title} {Rigorous qcd predictions for decays of
  $p$-wave quarkonia},\ }\href {https://doi.org/10.1103/PhysRevD.46.R1914}
  {\bibfield  {journal} {\bibinfo  {journal} {Phys. Rev. D}\ }\textbf {\bibinfo
  {volume} {46}},\ \bibinfo {pages} {R1914} (\bibinfo {year}
  {1992})}\BibitemShut {NoStop}%
\bibitem [{\citenamefont {Munz}(1996)}]{Munz}%
  \BibitemOpen
  \bibfield  {author} {\bibinfo {author} {\bibfnamefont {C.~R.}\ \bibnamefont
  {Munz}},\ }\bibfield  {title} {\bibinfo {title} {{Two photon decays of mesons
  in a relativistic quark model}},\ }\href
  {https://doi.org/10.1016/S0375-9474(96)00265-5} {\bibfield  {journal}
  {\bibinfo  {journal} {Nucl. Phys. A}\ }\textbf {\bibinfo {volume} {609}},\
  \bibinfo {pages} {364} (\bibinfo {year} {1996})},\ \Eprint
  {https://arxiv.org/abs/hep-ph/9601206} {arXiv:hep-ph/9601206} \BibitemShut
  {NoStop}%
\bibitem [{\citenamefont {Barbieri}\ \emph {et~al.}(1976)\citenamefont
  {Barbieri}, \citenamefont {Gatto},\ and\ \citenamefont
  {Kögerler}}]{BARBIERI1976183}%
  \BibitemOpen
  \bibfield  {author} {\bibinfo {author} {\bibfnamefont {R.}~\bibnamefont
  {Barbieri}}, \bibinfo {author} {\bibfnamefont {R.}~\bibnamefont {Gatto}},\
  and\ \bibinfo {author} {\bibfnamefont {R.}~\bibnamefont {Kögerler}},\
  }\bibfield  {title} {\bibinfo {title} {Calculation of the annihilation rate
  of p wave quark-antiquark bound states},\ }\href
  {https://doi.org/https://doi.org/10.1016/0370-2693(76)90419-6} {\bibfield
  {journal} {\bibinfo  {journal} {Physics Letters B}\ }\textbf {\bibinfo
  {volume} {60}},\ \bibinfo {pages} {183} (\bibinfo {year} {1976})}\BibitemShut
  {NoStop}%
\bibitem [{\citenamefont {Barbieri}\ \emph {et~al.}(1980)\citenamefont
  {Barbieri}, \citenamefont {Caffo}, \citenamefont {Gatto},\ and\ \citenamefont
  {Remiddi}}]{BARBIERI80}%
  \BibitemOpen
  \bibfield  {author} {\bibinfo {author} {\bibfnamefont {R.}~\bibnamefont
  {Barbieri}}, \bibinfo {author} {\bibfnamefont {M.}~\bibnamefont {Caffo}},
  \bibinfo {author} {\bibfnamefont {R.}~\bibnamefont {Gatto}},\ and\ \bibinfo
  {author} {\bibfnamefont {E.}~\bibnamefont {Remiddi}},\ }\bibfield  {title}
  {\bibinfo {title} {Strong qcd corrections to p-wave quarkonium decays},\
  }\href {https://doi.org/https://doi.org/10.1016/0370-2693(80)90407-4}
  {\bibfield  {journal} {\bibinfo  {journal} {Physics Letters B}\ }\textbf
  {\bibinfo {volume} {95}},\ \bibinfo {pages} {93} (\bibinfo {year}
  {1980})}\BibitemShut {NoStop}%
\bibitem [{\citenamefont {Barbieri}\ \emph {et~al.}(1981)\citenamefont
  {Barbieri}, \citenamefont {Caffo}, \citenamefont {Gatto},\ and\ \citenamefont
  {Remiddi}}]{BARBIERI81}%
  \BibitemOpen
  \bibfield  {author} {\bibinfo {author} {\bibfnamefont {R.}~\bibnamefont
  {Barbieri}}, \bibinfo {author} {\bibfnamefont {M.}~\bibnamefont {Caffo}},
  \bibinfo {author} {\bibfnamefont {R.}~\bibnamefont {Gatto}},\ and\ \bibinfo
  {author} {\bibfnamefont {E.}~\bibnamefont {Remiddi}},\ }\bibfield  {title}
  {\bibinfo {title} {Qcd corrections to p wave quarkonium decays},\ }\href
  {https://doi.org/https://doi.org/10.1016/0550-3213(81)90192-9} {\bibfield
  {journal} {\bibinfo  {journal} {Nuclear Physics B}\ }\textbf {\bibinfo
  {volume} {192}},\ \bibinfo {pages} {61} (\bibinfo {year} {1981})}\BibitemShut
  {NoStop}%
\bibitem [{\citenamefont {Ma}\ and\ \citenamefont {Wang}(2002)}]{MA2002233}%
  \BibitemOpen
  \bibfield  {author} {\bibinfo {author} {\bibfnamefont {J.}~\bibnamefont
  {Ma}}\ and\ \bibinfo {author} {\bibfnamefont {Q.}~\bibnamefont {Wang}},\
  }\bibfield  {title} {\bibinfo {title} {Corrections for two photon decays of
  ${\ensuremath{\chi}}_{c0}$ and ${\ensuremath{\chi}}_{c2}$ and color octet
  contributions},\ }\href
  {https://doi.org/https://doi.org/10.1016/S0370-2693(02)01937-8} {\bibfield
  {journal} {\bibinfo  {journal} {Physics Letters B}\ }\textbf {\bibinfo
  {volume} {537}},\ \bibinfo {pages} {233} (\bibinfo {year}
  {2002})}\BibitemShut {NoStop}%
\bibitem [{\citenamefont {Brambilla}\ \emph {et~al.}(2006)\citenamefont
  {Brambilla}, \citenamefont {Mereghetti},\ and\ \citenamefont
  {Vairo}}]{Brambilla}%
  \BibitemOpen
  \bibfield  {author} {\bibinfo {author} {\bibfnamefont {N.}~\bibnamefont
  {Brambilla}}, \bibinfo {author} {\bibfnamefont {E.}~\bibnamefont
  {Mereghetti}},\ and\ \bibinfo {author} {\bibfnamefont {A.}~\bibnamefont
  {Vairo}},\ }\bibfield  {title} {\bibinfo {title} {Electromagnetic quarkonium
  decays at order $v^7$},\ }\href
  {https://doi.org/10.1088/1126-6708/2006/08/039} {\bibfield  {journal}
  {\bibinfo  {journal} {Journal of High Energy Physics}\ }\textbf {\bibinfo
  {volume} {2006}},\ \bibinfo {pages} {039} (\bibinfo {year}
  {2006})}\BibitemShut {NoStop}%
\bibitem [{\citenamefont {Schuler}\ \emph {et~al.}(1998)\citenamefont
  {Schuler}, \citenamefont {Berends},\ and\ \citenamefont {{van
  Gulik}}}]{SCHULER}%
  \BibitemOpen
  \bibfield  {author} {\bibinfo {author} {\bibfnamefont {G.}~\bibnamefont
  {Schuler}}, \bibinfo {author} {\bibfnamefont {F.}~\bibnamefont {Berends}},\
  and\ \bibinfo {author} {\bibfnamefont {R.}~\bibnamefont {{van Gulik}}},\
  }\bibfield  {title} {\bibinfo {title} {Meson-photon transition form factors
  and resonance cross-sections in ${e}^{+}{e}^{\ensuremath{-}}$ collisions},\
  }\href {https://doi.org/https://doi.org/10.1016/S0550-3213(98)00128-X}
  {\bibfield  {journal} {\bibinfo  {journal} {Nuclear Physics B}\ }\textbf
  {\bibinfo {volume} {523}},\ \bibinfo {pages} {423} (\bibinfo {year}
  {1998})}\BibitemShut {NoStop}%
\bibitem [{\citenamefont {Sang}\ \emph {et~al.}(2016)\citenamefont {Sang},
  \citenamefont {Feng}, \citenamefont {Jia},\ and\ \citenamefont
  {Liang}}]{sang16}%
  \BibitemOpen
  \bibfield  {author} {\bibinfo {author} {\bibfnamefont {W.-L.}\ \bibnamefont
  {Sang}}, \bibinfo {author} {\bibfnamefont {F.}~\bibnamefont {Feng}}, \bibinfo
  {author} {\bibfnamefont {Y.}~\bibnamefont {Jia}},\ and\ \bibinfo {author}
  {\bibfnamefont {S.-R.}\ \bibnamefont {Liang}},\ }\bibfield  {title} {\bibinfo
  {title} {Next-to-next-to-leading-order qcd corrections to
  ${\ensuremath{\chi}}_{c0,2}\ensuremath{\rightarrow}\ensuremath{\gamma}\ensuremath{\gamma}$},\
  }\href {https://doi.org/10.1103/PhysRevD.94.111501} {\bibfield  {journal}
  {\bibinfo  {journal} {Phys. Rev. D}\ }\textbf {\bibinfo {volume} {94}},\
  \bibinfo {pages} {111501} (\bibinfo {year} {2016})}\BibitemShut {NoStop}%
\bibitem [{\citenamefont {Lansberg}\ and\ \citenamefont
  {Pham}(2009)}]{lansberg09}%
  \BibitemOpen
  \bibfield  {author} {\bibinfo {author} {\bibfnamefont {J.~P.}\ \bibnamefont
  {Lansberg}}\ and\ \bibinfo {author} {\bibfnamefont {T.~N.}\ \bibnamefont
  {Pham}},\ }\bibfield  {title} {\bibinfo {title} {Effective lagrangian for
  two-photon and two-gluon decays of $p$-wave heavy quarkonium
  ${\ensuremath{\chi}}_{c0,2}$ and ${\ensuremath{\chi}}_{b0,2}$ states},\
  }\href {https://doi.org/10.1103/PhysRevD.79.094016} {\bibfield  {journal}
  {\bibinfo  {journal} {Phys. Rev. D}\ }\textbf {\bibinfo {volume} {79}},\
  \bibinfo {pages} {094016} (\bibinfo {year} {2009})}\BibitemShut {NoStop}%
\bibitem [{\citenamefont {Chen}\ \emph {et~al.}(2017)\citenamefont {Chen},
  \citenamefont {Ding}, \citenamefont {Chang},\ and\ \citenamefont
  {Liu}}]{chen17}%
  \BibitemOpen
  \bibfield  {author} {\bibinfo {author} {\bibfnamefont {J.}~\bibnamefont
  {Chen}}, \bibinfo {author} {\bibfnamefont {M.}~\bibnamefont {Ding}}, \bibinfo
  {author} {\bibfnamefont {L.}~\bibnamefont {Chang}},\ and\ \bibinfo {author}
  {\bibfnamefont {Y.-x.}\ \bibnamefont {Liu}},\ }\bibfield  {title} {\bibinfo
  {title} {Two-photon transition form factor of $\overline{c}c$ quarkonia},\
  }\href {https://doi.org/10.1103/PhysRevD.95.016010} {\bibfield  {journal}
  {\bibinfo  {journal} {Phys. Rev. D}\ }\textbf {\bibinfo {volume} {95}},\
  \bibinfo {pages} {016010} (\bibinfo {year} {2017})}\BibitemShut {NoStop}%
\bibitem [{\citenamefont {Dudek}\ and\ \citenamefont
  {Edwards}(2006)}]{dudek06}%
  \BibitemOpen
  \bibfield  {author} {\bibinfo {author} {\bibfnamefont {J.~J.}\ \bibnamefont
  {Dudek}}\ and\ \bibinfo {author} {\bibfnamefont {R.~G.}\ \bibnamefont
  {Edwards}},\ }\bibfield  {title} {\bibinfo {title} {Two-photon decays of
  charmonia from lattice qcd},\ }\href
  {https://doi.org/10.1103/PhysRevLett.97.172001} {\bibfield  {journal}
  {\bibinfo  {journal} {Phys. Rev. Lett.}\ }\textbf {\bibinfo {volume} {97}},\
  \bibinfo {pages} {172001} (\bibinfo {year} {2006})}\BibitemShut {NoStop}%
\bibitem [{\citenamefont {Chen}\ \emph {et~al.}(2020)\citenamefont {Chen} \emph
  {et~al.}}]{Chen_2020}%
  \BibitemOpen
  \bibfield  {author} {\bibinfo {author} {\bibfnamefont {Y.}~\bibnamefont
  {Chen}} \emph {et~al.} (\bibinfo {collaboration} {CLQCD Collaboration}),\
  }\bibfield  {title} {\bibinfo {title} {Lattice study of two-photon decay
  widths for scalar and pseudo-scalar charmonium},\ }\href
  {https://doi.org/10.1088/1674-1137/44/8/083108} {\bibfield  {journal}
  {\bibinfo  {journal} {Chinese Physics C}\ }\textbf {\bibinfo {volume} {44}},\
  \bibinfo {pages} {083108} (\bibinfo {year} {2020})}\BibitemShut {NoStop}%
\bibitem [{\citenamefont {Crater}\ \emph {et~al.}(2006)\citenamefont {Crater},
  \citenamefont {Wong},\ and\ \citenamefont {Van~Alstine}}]{Crater}%
  \BibitemOpen
  \bibfield  {author} {\bibinfo {author} {\bibfnamefont {H.~W.}\ \bibnamefont
  {Crater}}, \bibinfo {author} {\bibfnamefont {C.-Y.}\ \bibnamefont {Wong}},\
  and\ \bibinfo {author} {\bibfnamefont {P.}~\bibnamefont {Van~Alstine}},\
  }\bibfield  {title} {\bibinfo {title} {Tests of two-body dirac equation wave
  functions in the decays of quarkonium and positronium into two photons},\
  }\href {https://doi.org/10.1103/PhysRevD.74.054028} {\bibfield  {journal}
  {\bibinfo  {journal} {Phys. Rev. D}\ }\textbf {\bibinfo {volume} {74}},\
  \bibinfo {pages} {054028} (\bibinfo {year} {2006})}\BibitemShut {NoStop}%
\bibitem [{\citenamefont {Wang}(2007)}]{WANG}%
  \BibitemOpen
  \bibfield  {author} {\bibinfo {author} {\bibfnamefont {G.-L.}\ \bibnamefont
  {Wang}},\ }\bibfield  {title} {\bibinfo {title} {Annihilation rate of heavy
  0++ p-wave quarkonium in relativistic salpeter method},\ }\href
  {https://doi.org/https://doi.org/10.1016/j.physletb.2007.08.017} {\bibfield
  {journal} {\bibinfo  {journal} {Physics Letters B}\ }\textbf {\bibinfo
  {volume} {653}},\ \bibinfo {pages} {206} (\bibinfo {year}
  {2007})}\BibitemShut {NoStop}%
\bibitem [{\citenamefont {Laverty}\ \emph {et~al.}(2011)\citenamefont
  {Laverty}, \citenamefont {Radford},\ and\ \citenamefont
  {Repko}}]{laverty2011gaga}%
  \BibitemOpen
  \bibfield  {author} {\bibinfo {author} {\bibfnamefont {J.~T.}\ \bibnamefont
  {Laverty}}, \bibinfo {author} {\bibfnamefont {S.~F.}\ \bibnamefont
  {Radford}},\ and\ \bibinfo {author} {\bibfnamefont {W.~W.}\ \bibnamefont
  {Repko}},\ }\href@noop {} {\bibinfo {title} {$\gamma\gamma$ and g g decay
  rates for equal mass heavy quarkonia}} (\bibinfo {year} {2011}),\ \Eprint
  {https://arxiv.org/abs/0901.3917} {arXiv:0901.3917 [hep-ph]} \BibitemShut
  {NoStop}%
\bibitem [{\citenamefont {Meng}\ \emph {et~al.}(2021)\citenamefont {Meng},
  \citenamefont {Feng}, \citenamefont {Liu}, \citenamefont {Wang},\ and\
  \citenamefont {Zou}}]{meng2021firstprinciple}%
  \BibitemOpen
  \bibfield  {author} {\bibinfo {author} {\bibfnamefont {Y.}~\bibnamefont
  {Meng}}, \bibinfo {author} {\bibfnamefont {X.}~\bibnamefont {Feng}}, \bibinfo
  {author} {\bibfnamefont {C.}~\bibnamefont {Liu}}, \bibinfo {author}
  {\bibfnamefont {T.}~\bibnamefont {Wang}},\ and\ \bibinfo {author}
  {\bibfnamefont {Z.}~\bibnamefont {Zou}},\ }\bibfield  {title} {\bibinfo
  {title} {First-principle calculation of $\eta_c\rightarrow 2\gamma$ decay
  width from lattice qcd},\ }\href@noop {} {\  (\bibinfo {year} {2021})},\
  \Eprint {https://arxiv.org/abs/2109.09381} {arXiv:2109.09381 [hep-lat]}
  \BibitemShut {NoStop}%
\bibitem [{\citenamefont {Ji}\ and\ \citenamefont {Jung}(2001)}]{Ji}%
  \BibitemOpen
  \bibfield  {author} {\bibinfo {author} {\bibfnamefont {X.}~\bibnamefont
  {Ji}}\ and\ \bibinfo {author} {\bibfnamefont {C.}~\bibnamefont {Jung}},\
  }\bibfield  {title} {\bibinfo {title} {Studying hadronic structure of the
  photon in lattice qcd},\ }\href {https://doi.org/10.1103/PhysRevLett.86.208}
  {\bibfield  {journal} {\bibinfo  {journal} {Phys. Rev. Lett.}\ }\textbf
  {\bibinfo {volume} {86}},\ \bibinfo {pages} {208} (\bibinfo {year}
  {2001})}\BibitemShut {NoStop}%
\bibitem [{\citenamefont {McNeile}\ and\ \citenamefont
  {Michael}(2004)}]{ukqcd}%
  \BibitemOpen
  \bibfield  {author} {\bibinfo {author} {\bibfnamefont {C.}~\bibnamefont
  {McNeile}}\ and\ \bibinfo {author} {\bibfnamefont {C.}~\bibnamefont
  {Michael}} (\bibinfo {collaboration} {UKQCD Collaboration}),\ }\bibfield
  {title} {\bibinfo {title} {Estimate of the flavor singlet contributions to
  the hyperfine splitting in charmonium},\ }\href
  {https://doi.org/10.1103/PhysRevD.70.034506} {\bibfield  {journal} {\bibinfo
  {journal} {Phys. Rev. D}\ }\textbf {\bibinfo {volume} {70}},\ \bibinfo
  {pages} {034506} (\bibinfo {year} {2004})}\BibitemShut {NoStop}%
\bibitem [{\citenamefont {Forcrand}\ \emph {et~al.}(2004)\citenamefont
  {Forcrand} \emph {et~al.}}]{taro}%
  \BibitemOpen
  \bibfield  {author} {\bibinfo {author} {\bibfnamefont {P.}~\bibnamefont
  {Forcrand}} \emph {et~al.} (\bibinfo {collaboration} {The QCD-TARO
  Collaboration}),\ }\bibfield  {title} {\bibinfo {title} {Contribution of
  disconnected diagrams to the hyperfine splitting of charmonium},\ }\href
  {https://doi.org/10.1088/1126-6708/2004/08/004} {\bibfield  {journal}
  {\bibinfo  {journal} {JHEP}\ }\textbf {\bibinfo {volume} {2004}}\bibinfo
  {number} { (08)},\ \bibinfo {pages} {004}}\BibitemShut {NoStop}%
\bibitem [{\citenamefont {Shultz}\ \emph {et~al.}(2015)\citenamefont {Shultz},
  \citenamefont {Dudek},\ and\ \citenamefont {Edwards}}]{shultz15}%
  \BibitemOpen
\bibfield  {number} {  }\bibfield  {author} {\bibinfo {author} {\bibfnamefont
  {C.~J.}\ \bibnamefont {Shultz}}, \bibinfo {author} {\bibfnamefont {J.~J.}\
  \bibnamefont {Dudek}},\ and\ \bibinfo {author} {\bibfnamefont {R.~G.}\
  \bibnamefont {Edwards}} (\bibinfo {collaboration} {for the Hadron Spectrum
  Collaboration}),\ }\bibfield  {title} {\bibinfo {title} {Excited meson
  radiative transitions from lattice qcd using variationally optimized
  operators},\ }\href {https://doi.org/10.1103/PhysRevD.91.114501} {\bibfield
  {journal} {\bibinfo  {journal} {Phys. Rev. D}\ }\textbf {\bibinfo {volume}
  {91}},\ \bibinfo {pages} {114501} (\bibinfo {year} {2015})}\BibitemShut
  {NoStop}%
\bibitem [{\citenamefont {Boucaud}\ \emph {et~al.}(2008)\citenamefont {Boucaud}
  \emph {et~al.}}]{BOUCAUD2008695}%
  \BibitemOpen
  \bibfield  {author} {\bibinfo {author} {\bibfnamefont {P.}~\bibnamefont
  {Boucaud}} \emph {et~al.},\ }\bibfield  {title} {\bibinfo {title} {Dynamical
  twisted mass fermions with light quarks: simulation and analysis details},\
  }\href {https://doi.org/https://doi.org/10.1016/j.cpc.2008.06.013} {\bibfield
   {journal} {\bibinfo  {journal} {Comput. Phys. Commun.}\ }\textbf {\bibinfo
  {volume} {179}},\ \bibinfo {pages} {695} (\bibinfo {year}
  {2008})}\BibitemShut {NoStop}%
\bibitem [{\citenamefont {Blossier}\ \emph {et~al.}(2010)\citenamefont
  {Blossier}, \citenamefont {Dimopoulos}, \citenamefont {Frezzotti},
  \citenamefont {Lubicz}, \citenamefont {Petschlies}, \citenamefont
  {Sanfilippo}, \citenamefont {Simula},\ and\ \citenamefont
  {Tarantino}}]{tw2010}%
  \BibitemOpen
  \bibfield  {author} {\bibinfo {author} {\bibfnamefont {B.}~\bibnamefont
  {Blossier}}, \bibinfo {author} {\bibfnamefont {P.}~\bibnamefont
  {Dimopoulos}}, \bibinfo {author} {\bibfnamefont {R.}~\bibnamefont
  {Frezzotti}}, \bibinfo {author} {\bibfnamefont {V.}~\bibnamefont {Lubicz}},
  \bibinfo {author} {\bibfnamefont {M.}~\bibnamefont {Petschlies}}, \bibinfo
  {author} {\bibfnamefont {F.}~\bibnamefont {Sanfilippo}}, \bibinfo {author}
  {\bibfnamefont {S.}~\bibnamefont {Simula}},\ and\ \bibinfo {author}
  {\bibfnamefont {C.}~\bibnamefont {Tarantino}},\ }\bibfield  {title} {\bibinfo
  {title} {Average up/down, strange, and charm quark masses with ${N}_{f}=2$
  twisted-mass lattice qcd},\ }\href
  {https://doi.org/10.1103/PhysRevD.82.114513} {\bibfield  {journal} {\bibinfo
  {journal} {Phys. Rev. D}\ }\textbf {\bibinfo {volume} {82}},\ \bibinfo
  {pages} {114513} (\bibinfo {year} {2010})}\BibitemShut {NoStop}%
\bibitem [{\citenamefont {Zyla}\ \emph {et~al.}(2020)\citenamefont {Zyla} \emph
  {et~al.}}]{Zyla:2020zbs}%
  \BibitemOpen
  \bibfield  {author} {\bibinfo {author} {\bibfnamefont {P.}~\bibnamefont
  {Zyla}} \emph {et~al.} (\bibinfo {collaboration} {Particle Data Group}),\
  }\bibfield  {title} {\bibinfo {title} {{Review of Particle Physics}},\ }\href
  {https://doi.org/10.1093/ptep/ptaa104} {\bibfield  {journal} {\bibinfo
  {journal} {PTEP}\ }\textbf {\bibinfo {volume} {2020}},\ \bibinfo {pages}
  {083C01} (\bibinfo {year} {2020})}\BibitemShut {NoStop}%
\bibitem [{\citenamefont {Albanese}\ \emph {et~al.}(1987)\citenamefont
  {Albanese} \emph {et~al.}}]{ape}%
  \BibitemOpen
  \bibfield  {author} {\bibinfo {author} {\bibfnamefont {M.}~\bibnamefont
  {Albanese}} \emph {et~al.} (\bibinfo {collaboration} {APE Collaboration}),\
  }\bibfield  {title} {\bibinfo {title} {Glueball masses and string tension in
  lattice qcd},\ }\href
  {https://doi.org/https://doi.org/10.1016/0370-2693(87)91160-9} {\bibfield
  {journal} {\bibinfo  {journal} {Phys. Lett. B}\ }\textbf {\bibinfo {volume}
  {192}},\ \bibinfo {pages} {163} (\bibinfo {year} {1987})}\BibitemShut
  {NoStop}%
\bibitem [{\citenamefont {Gusken}(1990)}]{Gusken:1989qx}%
  \BibitemOpen
  \bibfield  {author} {\bibinfo {author} {\bibfnamefont {S.}~\bibnamefont
  {Gusken}},\ }\bibfield  {title} {\bibinfo {title} {{A Study of smearing
  techniques for hadron correlation functions}},\ }\href
  {https://doi.org/10.1016/0920-5632(90)90273-W} {\bibfield  {journal}
  {\bibinfo  {journal} {Nucl. Phys. B Proc. Suppl.}\ }\textbf {\bibinfo
  {volume} {17}},\ \bibinfo {pages} {361} (\bibinfo {year} {1990})}\BibitemShut
  {NoStop}%
\bibitem [{\citenamefont {Bali}\ \emph {et~al.}(2011)\citenamefont {Bali},
  \citenamefont {Collins} \emph {et~al.}}]{bali}%
  \BibitemOpen
  \bibfield  {author} {\bibinfo {author} {\bibfnamefont {G.}~\bibnamefont
  {Bali}}, \bibinfo {author} {\bibfnamefont {S.}~\bibnamefont {Collins}}, \emph
  {et~al.},\ }\bibfield  {title} {\bibinfo {title} {Spectra of heavy-light and
  heavy-heavy mesons containing charm quarks, including higher spin states for
  $n_f=2+ 1$},\ }\href@noop {} {\bibfield  {journal} {\bibinfo  {journal} {PoS
  \textbf{LATTICE2011}}\ ,\ \bibinfo {pages} {135}} (\bibinfo {year} {2011})},\
  \Eprint {https://arxiv.org/abs/1108.6147} {arXiv:1108.6147 [hep-lat]}
  \BibitemShut {NoStop}%
\end{thebibliography}
%apsrev4-2.bst 2019-01-14 (MD) hand-edited version of apsrev4-1.bst
%Control: key (0)
%Control: author (8) initials jnrlst
%Control: editor formatted (1) identically to author
%Control: production of article title (0) allowed
%Control: page (0) single
%Control: year (1) truncated
%Control: production of eprint (0) enabled
%

\end{document}